\documentclass[preprint,amsmath,amssymb, aps, prd,groupedaddress, superscriptaddress, nofootinbib]{revtex4-2}

\usepackage{mathtools}
\usepackage{graphicx}
\usepackage{dcolumn}   
\usepackage{bm}        
\usepackage{verbatim}   
\usepackage{lineno}
\usepackage{lineno, booktabs, setspace}
\usepackage{wrapfig}
\usepackage{float, array}
\usepackage{amsmath}
\usepackage{amssymb}
\usepackage{verbatim, url}
\usepackage{enumerate}
\usepackage{bm}
\usepackage{notes2bib}
\usepackage{lpic}
\usepackage{pgfplots}
\usepackage[latin1]{inputenc}
\usepackage[extra]{tipa}
\usepackage{accents}
\usepackage{epstopdf}
\usepackage{appendix}
\usepackage{amsthm} 
\usepackage{booktabs}

\newcommand{\bra}[1]{\langle #1|}
\newcommand{\ket}[1]{|#1\rangle}

\newcommand{\nn}{\nonumber}
\newcommand{\f}[2] {\frac{#1}{#2}}

\newcommand{\beq}{\begin{equation}}
\newcommand{\eeq}{\end{equation}}
\newcommand{\beqn}{\begin{eqnarray}}
\newcommand{\eeqn}{\end{eqnarray}}
\newcommand{\la}{\left\langle}
\newcommand{\ra}{\right\rangle}
\DeclarePairedDelimiter\abs{\lvert}{\rvert}%
\DeclarePairedDelimiter\norm{\lVert}{\rVert}%
\makeatletter
\let\oldabs\abs
\def\abs{\@ifstar{\oldabs}{\oldabs*}}
\let\oldnorm\norm
\def\norm{\@ifstar{\oldnorm}{\oldnorm*}}
\makeatother

\begin{document}

\count\footins = 800

\title{\boldmath Eigenstate Thermalization in the Two-Site SYK and SYK Chain Models}
	
\author{Seyyed M.H. Halataei}
\affiliation{Department of Physics, Shahid Beheshti University, Tehran, Iran}

\date{\today}

\begin{abstract}
	A recent study of R\'enyi entanglement entropy in the SYK chain of Majorana fermions suggested that the model does not rapidly thermalize, despite being maximally chaotic. In this work, I examine the Eigenstate Thermalization Hypothesis (ETH) for both the SYK chain and the two-site SYK models using exact diagonalization. I show that single realizations of both models approximately satisfy ETH conditions, while ensemble averages strictly satisfy ETH. Therefore, I conclude that the finite-size SYK chain and two-site SYK models can rapidly thermalize with respect to generic few-body operators through the ETH mechanism. This suggests that the subthermal behavior observed in previous studies of R\'enyi entanglement entropy does not manifest in finite-size systems, and that these systems can thermalize rapidly via their light modes. It also indicates that the proposed gravitational dual may undergo rapid thermalization. 
\end{abstract}

\maketitle
	
\newpage

\section{Introduction}

Chaotic quantum many-body systems are generally expected to thermalize rapidly, even when isolated from their environments. The eigenstate thermalization hypothesis (ETH) provides a theoretical foundation for this phenomenon \cite{Deutsch1991, Srednicki1994, Srednicki1996}. ETH asserts that a closed quantum system initialized in a pure state far from equilibrium can thermalize with respect to a set of observables, provided those observables satisfy certain conditions. In such cases, the expectation values of the observables evolve toward thermal values and remain close to them for most of the time, with only exponentially small fluctuations.

However, recent investigations of Rényi entanglement entropy in maximally chaotic systems have raised questions about the universality and speed of thermalization. In particular, evidence for slow thermalization has been reported in a one-dimensional generalization of the Sachdev-Ye-Kitaev (SYK) model, known as the SYK chain model, including its special case with two sites \cite{Gu2017b, Gu2017}.

In this work, I study thermalization in the two-site SYK and SYK chain models from the perspective of ETH. The SYK chain is a one-dimensional lattice model with quenched disorder, where each site hosts $N$ Majorana fermions with quartic random interactions \cite{Gu2017}. Neighboring sites interact via 2-2 couplings between Majorana fermions. This model retains several key features of the original zero-dimensional SYK model \cite{Kitaev2015, Maldacena2016}, such as local criticality, extensive zero-temperature entropy, and maximal chaos. It also exhibits additional properties, including diffusive energy transport and butterfly velocity for the propagation of chaos in space \cite{Gu2017}. From a condensed matter perspective, the SYK chain is a rare example of a solvable, strongly correlated, chaotic lattice model. It provides a valuable platform for studying thermalization, entanglement propagation, many-body localization, quantum phase transitions, and dissipative transport.

From the perspective of holographic duality, the SYK chain is conjectured to be dual to an incoherent black hole \cite{Gu2017b}. Since the isolated SYK chain evolves unitarily, understanding its thermalization dynamics may shed light on the formation and evaporation of such a black hole, and contribute to ongoing discussions surrounding the black hole information paradox \cite{Hawking1976}. \footnote{For recent progress on resolving the information paradox and related challenges, see \cite{Penington2020, Almheiri2019a, Almheiri2020a, Almheiri2019b, Almheiri2020b, Almheiri2020c, Alishahiha2021, Bousso2020, Saad2021}.}

In this paper, I examine the validity of ETH in the two-site SYK and SYK chain models by studying two conventional, generic, simple, few-body operators. I analyze both individual disorder realizations and ensemble-averaged data. Through exact diagonalization, I find that individual realizations exhibit what I term ``approximate ETH behavior," while ensemble-averaged results exhibit ETH behavior more precisely. These findings suggest that the SYK chain and two-site SYK, and by extension its putative holographic duals, can indeed thermalize rapidly via the ETH mechanism. I discuss the apparent discrepancy with the conclusions of the Rényi entropy study \cite{Gu2017b} in the final section of the paper.

Previous studies have demonstrated ETH in various zero-dimensional SYK-like models, including the original SYK model \cite{Hunter2018}, the complex SYK model \cite{Sonner2017}, and the supersymmetric SYK model \cite{Hunter2018}. These models feature all-to-all interactions without spatial structure. The influence of spatial locality and the absence of all-to-all couplings on ETH behavior has been noted as an open question \cite{Hunter2018}.  The results presented in this work suggest that, within the analytically tractable regime considered, the presence of spatial locality and the absence of all-to-all interactions do not obstruct the emergence of ETH-like thermalization in the SYK chain and two-site SYK.

Furthermore, by analyzing the off-diagonal matrix elements of the operators, I demonstrate that the SYK chain and two-site SYK exhibits behavior consistent with random matrix theory (RMT) up to a characteristic energy scale, beyond which deviations from RMT emerge. Drawing an analogy with disordered conductors, I identify this scale as the Thouless energy, $E_T$, and show that it increases with the inter-site coupling strength $J_1$. In particular, stronger coupling between adjacent SYK sites extends the energy window over which the system displays RMT-like behavior.\footnote{For discussions of RMT behavior and the Thouless energy in the SYK and supersymmetric SYK models, see \cite{Garcia2016, Garcia2018}.}

The remainder of the paper is organized as follows. In Section \ref{Review}, I review the eigenstate thermalization hypothesis (ETH) and the two-site SYK and SYK chain models. I introduce the concept of \textit{approximate ETH behavior} and discuss the Thouless energy and diffusion constant of the SYK chain model. In Section \ref{Setup}, I describe the computational setup, including the numerical construction of the two-site SYK and SYK chain models, the exploration of various coupling regimes, the construction of few-body operators, and the determination of their thermal values along with a representative entropy for ETH comparisons. In Section~\ref{NumericalChecks-Single}, I numerically verify ETH for single realizations of the two-site SYK and SYK chain models and demonstrate that ETH is approximately satisfied. In Section~\ref{NumericalChecks-Ensemble}, I verify ETH for the ensemble-averaged models, showing that they satisfy ETH precisely. I also identify the Thouless energy and examine the relation between the diffusion constant and the coupling strengths. Finally, in Section \ref{con}, I conclude with a summary of the results, a discussion of their implications and future directions, and a comparison with the study of R\'enyi entanglement entropy.

\section{Overview} \label{Review}

\subsection{Eigenstate Thermalization Hypothesis} \label{OverviewETH}
The eigenstate thermalization hypothesis (ETH) explains how isolated quantum systems in pure states can exhibit thermal behavior despite evolving unitarily \cite{Deutsch1991, Srednicki1994, Srednicki1996}. In this subsection, I review the basic structure of the hypothesis.

Consider an isolated quantum system with finitely many degrees of freedom, $\infty > \mathcal{N} \gg 1$, and a non-degenerate, discrete energy spectrum $\{E_n\}$. Let $H$ denote the Hamiltonian of the system, with energy eigenstates ${\ket{n}}$. For any pure state $\ket{\psi}$ of the system, the quantum average (i.e., expectation value) of energy is given by

\beq
	\la H \ra_Q=\bra{\psi} H \ket{\psi}
\eeq
	and the quantum energy uncertainty is defined as
\beq
	\Delta H_Q = \left[ \bra{\psi} H^2 \ket{\psi} - \bra{\psi} H \ket{\psi}^2 \right]^{1/2}.
\eeq
ETH is applicable to pure states for which the quantum energy uncertainty is much smaller than the average energy:
\beq \label{DEEbar}
	\Delta H_Q \ll \la H \ra_Q. 
\eeq
It is sufficient for the energy uncertainty to scale as $\Delta H_Q \sim \langle H \rangle_Q / \mathcal{N}^{\nu}$ for some $\nu > 0$.

For later use, we associate to each pure state $\ket{\psi}$ an artificial temperature $1/\beta_\psi$ such that the expectation value of the Hamiltonian in pure state $\ket{\psi}$ and in thermal state $e^{-\beta_\psi H}$ are equal:
\beq \label{betapsi}
	\la H \ra_Q =  \bra{\psi} H \ket{\psi}= \f{\text{Tr } e^{-\beta_\psi H} H}{\text{Tr } e^{-\beta_\psi H}}= \la H \ra_{th}. 
\eeq
Equation above implicitly defines the parameter $\beta_\psi$, such that the average energy of the pure state $\ket{\psi}$ matches that of the corresponding thermal state $e^{-\beta_\psi H}$.

Next, consider a finite collection of generic observables $\left\{ A^{\alpha} \right\}_{\alpha=1}^{\Gamma}$, where the number of observables $\Gamma$ is much smaller than the dimension of the Hilbert space, i.e., $\Gamma \ll \text{dim}(\mathcal{H})$. These observables are considered generic in the sense that they correspond either to physically measurable quantities or to non-extensive operators acting on only a few degrees of freedom--for instance, the single-site number operator or two-site hopping operator.

	For any state $\ket{\psi}$, the set $\left\{ A^{\alpha} \right\}_{\alpha=1}^{\Gamma}$ are chosen such that the quantum average of the observables 
\beq
\la A^{\alpha} \ra_Q=\bra{\psi} A^{\alpha} \ket{\psi}
\eeq
	are far from their thermal averages at temperature $1/\beta_\psi$
\beq
	\la A^{\alpha} \ra_{th} = \f{\text{Tr } e^{-\beta_\psi H} A^{\alpha}}{\text{Tr } e^{-\beta_\psi H}}.
\eeq

Thus, as the system starts from state $\ket{\psi}$, the observables start from non-equilibrium initial expectation values $\la A^{\alpha}(0) \ra_Q  \ne \la A^{\alpha} \ra_{th}$ (see Fig. 1).

Expanding the pure state in terms of the energy eigenkets,
\beq
	\ket{\psi} = \sum_n c_n \ \ket{n},
\eeq
the time-evolved state becomes
\beq
	\ket{\psi(t)} = \sum_n c_n \ e^{-i E_n t/\hbar} \ \ket{n}.
\eeq
Accordingly, the expectation values of the observables evolve as
\beq \label{At}
	\la A^{\alpha}(t) \ra_Q =\bra{\psi(t)} A^{\alpha} \ket{\psi(t)} = \sum_{m,n} c_m^* c_n \ A^{\alpha}_{mn}\  e^{i (E_m-E_n)t /\hbar}
\eeq
	where
\beq
	A^{\alpha}_{mn} = \bra{m} A^{\alpha} \ket{n}
\eeq
	denotes the matrix element of the observable $A^{\alpha}$ in the energy eigenbasis. 
	
	Having established the framework, we now state the ETH.
	
	ETH asserts that, under the assumptions and setup above, if the observables $\left\{ A^{\alpha} \right\}_{\alpha=1}^{\Gamma}$ satisfy the following two conditions, their time-evolved expectation values $\langle A^{\alpha}(t) \rangle_Q$ will relax to thermal equilibrium values $\langle A^{\alpha} \rangle_{th}$ and remain close to them for most of the time (Fig. \ref{fig:eth-pic}):
		
	\begin{figure}
		\includegraphics[width=0.9\linewidth]{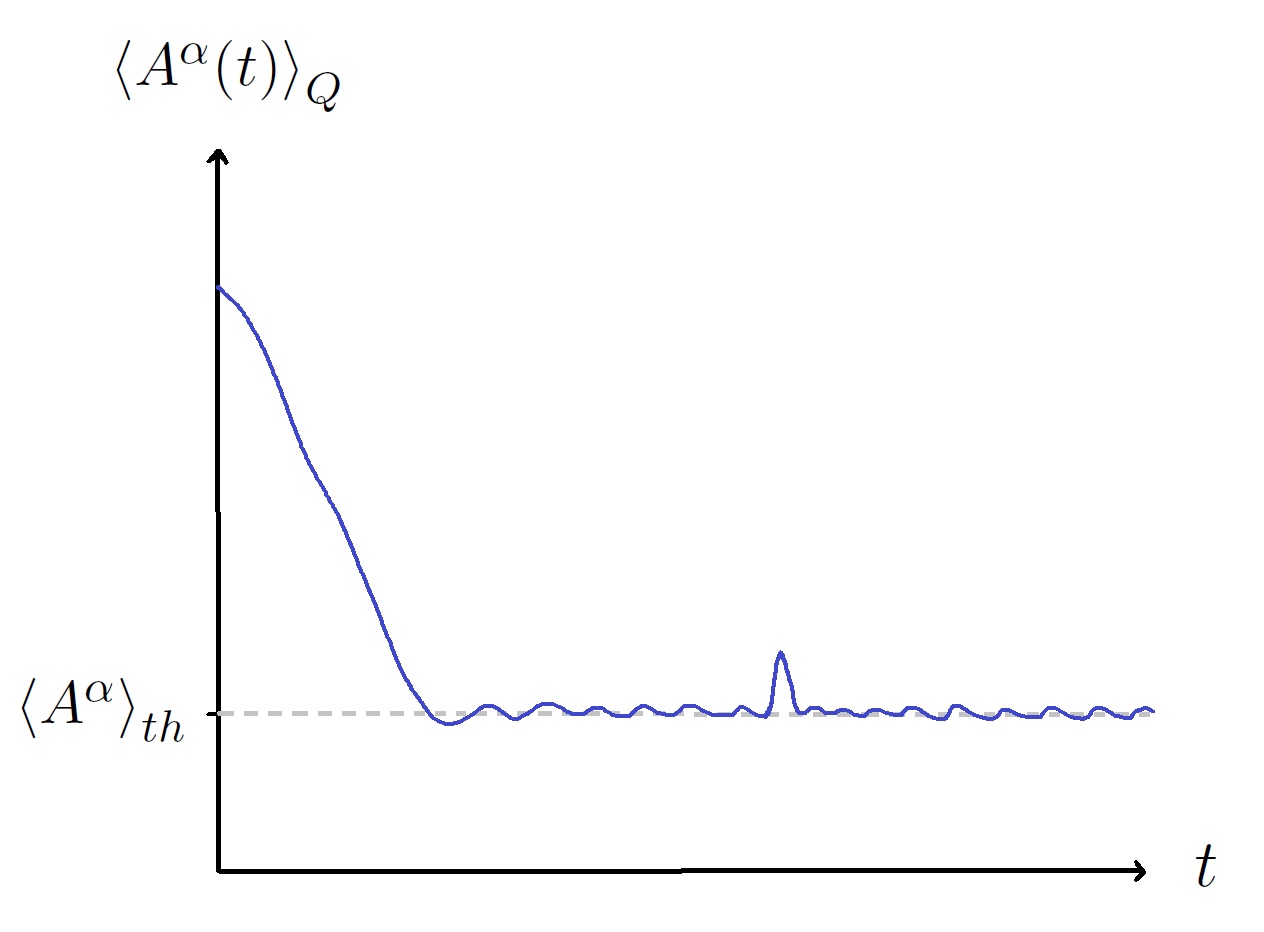}
		\caption{Eigenstate Thermalization Hypothesis. If ETH holds for a set of observables $\left\{ A^{\alpha} \right\}_{\alpha=1}^{\Gamma}$, then those observables thermalize under unitary evolution. They are chosen such that their quantum expectation values in the initial pure state lie far from equilibrium. ETH asserts that these expectation values $\langle A^\alpha (t) \rangle_Q$ relax to their thermal values $\langle A^\alpha \rangle_{\mathrm{th}}$ and remain close to them for most of the time. Fluctuations around equilibrium are exponentially suppressed, though rare large deviations may occur.}
		\label{fig:eth-pic}
	\end{figure}
	
The two conditions of ETH are as follows: For all $n$, $m$, and $1 \le \alpha \le \Gamma$, ETH requires:
	\beqn
	\label{ETH1} A^\alpha_{nn} &=& \mathcal{A}^\alpha(E_n), \\
	\label{ETH2} \abs{A^\alpha_{nm}} &\sim& e^{-S/2}
	\eeqn
	where $\mathcal{A}^\alpha$ is a smooth function of energy and $S$ is the entropy of the system, which scales extensively with the number of degrees of freedom, $S \sim \mathcal{N}$. In words, ETH requires that the diagonal matrix elements $A^\alpha_{nn}$ follow a smooth energy-dependent curve, and the off-diagonal elements be exponentially suppressed.	
	
	When these conditions hold, ETH implies thermalization with respect to the observables  $\left\{ A^{\alpha} \right\}_{\alpha=1}^{\Gamma}$: each observable relaxes to its thermal value and fluctuates only slightly around it for most of the time. The fluctuations are exponentially suppressed, though rare large deviations can occur.\footnote{In some literature, the conditions of ETH are expressed in terms of the following equation:
	\beq
	\label{ETH3} 
	A_{mn}^\alpha = \mathcal{A}^\alpha(E_n) \delta_{mn} + e^{-S/2} f^\alpha(\overline{E},\omega) R_{mn}
	\eeq
		where \( \overline{E} = (E_n + E_m)/2 \), \( \omega = E_m - E_n \), \( f^\alpha \) is a smooth function, and \( R_{mn} \) is a Gaussian random variable with zero mean and unit variance. The key difference between conditions \eqref{ETH1} and \eqref{ETH2}, as originally introduced by the founders of ETH \cite{Deutsch1991, Srednicki1994, Srednicki1996}, and the form in equation \eqref{ETH3} is the inclusion of the random variable \( R_{mn} \) in the off-diagonal matrix elements. However, it is not generally necessary for the off-diagonal elements to be randomly distributed for ETH to hold and for the closed system to reach thermal equilibrium. While quantized chaotic systems satisfy equation \eqref{ETH3}, the specific distribution (e.g., Gaussian) of the off-diagonal matrix elements is not required for thermalization. Therefore, we do not consider the distribution of off-diagonal elements in this work.}
	
To demonstrate thermalization, ETH shows that the infinite time average of the observables
\beq
\overline{A^\alpha} = \lim\limits_{T \to \infty} \f{1}{T} \int_{0}^{\infty} \la A^\alpha(t) \ra dt
\eeq
are approximately equal to the smooth function $\mathcal{A}^\alpha(E_n)$, given that the diagonal matrix elements satisfy Eq. \eqref{ETH1}. This leads to the result that the infinite time average values are close to the thermal and microcanonical average values of the observables:
\beq \label{result11}
\overline{A^\alpha} \simeq  \mathcal{A}^{\alpha}(E_n) = A^\alpha_{nn} \simeq  \la A^\alpha \ra_{th} \simeq \la A^\alpha \ra_{micro}.
\eeq
The approximations are valid up to $\mathcal{O}(1/S)$ and $\mathcal{O}(\Delta H ^2)$ \cite{Srednicki1994,Srednicki1996}. 

Furthermore, ETH shows that the infinite time average of the fluctuations around the thermal values are suppressed by the off digonal matrix elements as follows:
\beq \label{fluc}
\overline{\abs{\la A^\alpha(t) \ra_Q - \overline{A^\alpha}}^2} = \sum_{m \ne n} \abs{c_n}^2 \abs{c_m}^2 \abs{A^\alpha_{nm}}^2.
\eeq
If the off-digonal matrix elements satisfy Eq. \eqref{ETH2}, the right hand side of the above equation will be exponentially small:
\beq \label{flucSmall}
\overline{\abs{\la A^\alpha(t) \ra_Q - \overline{A^\alpha}}^2} \sim e^{-S}. 
\eeq
This occurs because $\abs{c_n}^2$ and $\abs{c_m}^2$ sum to unity.

Eq. \eqref{result11} demonstrates that the infinite time average values of the observables coincide with their thermal averages, while Eq. \eqref{flucSmall} guarantees that fluctuations around these thermal average values are exponentially small for most of the time.

This is ETH in its most rigorous form. However, in certain cases, not all the conditions of ETH are simultaneously satisfied. Nevertheless, the general conclusions of ETH regarding thermalization of observables still hold. For instance, in systems such as the single realization of the SYK chain model, as discussed in Sec. \ref{NumericalChecks-Single}, Eq. \eqref{ETH2} may not be satisfied, and the off-diagonal matrix elements are only sub-exponentially small, rather than exponentially small. Nonetheless, since their maximum absolute values are still small, one can use Eq. \eqref{fluc} to obtain:
\beq
\overline{\abs{\la A^\alpha(t) \ra_Q - \overline{A^\alpha}}^2} \leqslant \max_{n \ne m} \ \abs{A_{nm}}^2.
\eeq
and conclude that the fluctuations remain small, though not exponentially small. 

In such cases, we say that ETH is \textit{approximately} satisfied. As long as the off-diagonal matrix elements are one or more orders of magnitude smaller than the diagonal elements, significant deviations from the thermal average values do not occur during typical times, allowing us to assume that thermalization is achieved in the system.

Another point concerning the conditions of ETH is inequality \eqref{DEEbar}, which requires the quantum fluctuation of energy to be much smaller than the quantum average energy. This inequality is used in the proof of ETH to demonstrate Eq. \eqref{result11} for the diagonal matrix elements. If one can directly check the satisfaction of Eq. \eqref{result11}, the validity of inequality \eqref{DEEbar} does not need to be verified. This is what I do in Secs. \ref{NumericalChecks-Single}-\ref{NumericalChecks-Ensemble}. I directly check Eq. \eqref{result11} both for single realizations of the SYK chain and two-site SYK models and for the ensemble-averaged theories and present the results there.

\subsection{SYK Chain Model}
	The SYK chain model \cite{Gu2017} is a generalization of the original SYK model \cite{Kitaev2015, Maldacena2016}. In this subsection, I first provide a brief overview of the original SYK model and then describe the SYK chain model.
	
	The original SYK model is an ensemble-averaged theory \cite{Kitaev2015, Maldacena2016}. Each single realization of the ensemble consists of $N$ Majorana fermions with quartic random all-to-all interactions. The Hamiltonian for each realization is given by
	\beq
	H = \sum_{1\leqslant i < j < k < l \leqslant N} J_{ijkl} \ \chi_i \chi_j \chi_k \chi_l 
	\eeq
	where the Majorana fermion operators satisfy the Clifford algebra
	\beq
	\{ \chi_i, \chi_j \} = \delta_{ij}
	\eeq
	and the couplings $\{ J_{ijkl} \}$ are independent random real numbers drawn from a Gaussian distribution with zero mean 
	\beq
	\overline{J_{ijkl}} = 0, 
	\eeq
	and variance
	\beq
	\overline{J_{ijkl}^2} = \f{3! J^2}{N^3}.
	\eeq
	Here, $J$ sets the average strength of the couplings. 
	
	The SYK model is solvable in the large-$N$ limit. It exhibits many interesting properties at strong coupling $N \gg \beta J \gg 1$, such as extensive zero-temperature entropy, local criticality (with power law correlations in time), and maximal chaos. In this limit, the averaged theory also exhibits holographic behavior and demonstrates connections to nearly $AdS_2$ gravity. However, the model is (0+1)-dimensional and lacks spatial locality. Consequently, some expected properties of higher-dimensional systems, such as diffusion and the butterfly effect, cannot be realized in the original model.
	
\begin{figure} \includegraphics[width=\linewidth]{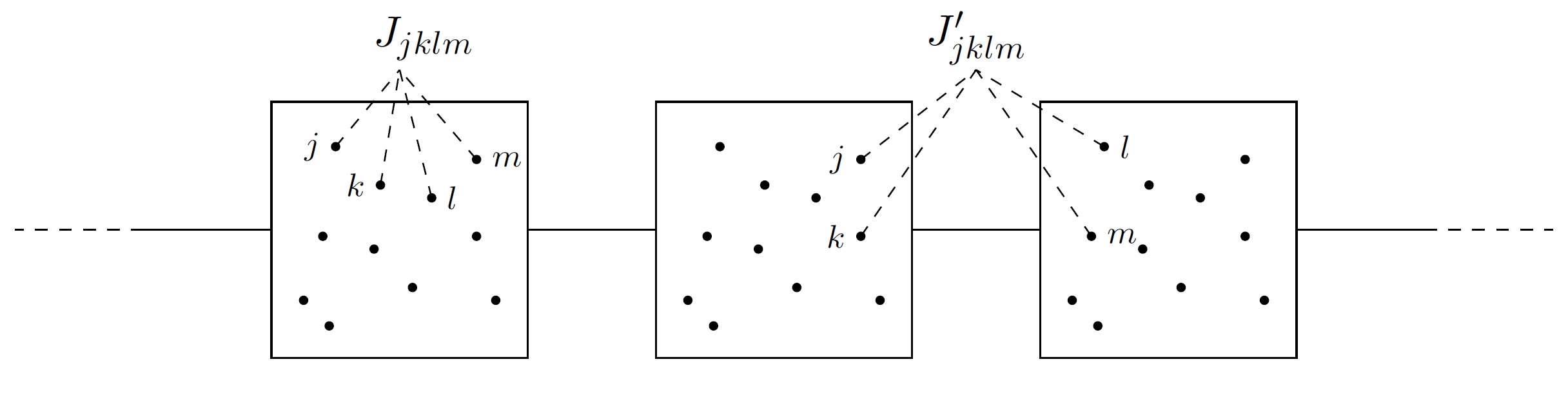} \caption{\label{fig:syk-chain} SYK chain model. The model consists of $M$ sites with periodic boundary conditions. Each site contains $N$ Majorana fermions with SYK interactions. Nearest neighboring sites interact via four-fermion interactions with two fermions from each site. (Graphic adapted from \cite{Gu2017})} \end{figure}

	A higher-dimensional lattice generalization of the SYK model with spatial locality was proposed by Gu et al. \cite{Gu2017}. In $(1+1)$-dimensions, the model consists of a chain of SYK sites (also known as links), referred to as the SYK chain model, as illustrated in Fig. \ref{fig:syk-chain}.
	
	In this paper, I focus on the SYK chain model and its special case with two sites. Like the original SYK model, the SYK chain model is an ensemble-averaged theory. The Hamiltonian for a single realization of the model is given by
	\beqn \label{HChain}
	\nn H &=& \sum_{x=1}^M \ \sum_{1 \leqslant i < j < k < l \leqslant N} J_{ijkl,x} \ \chi_{i,x} \chi_{j,x} \chi_{k,x} \chi_{l,x} \\ 
	&+& \sum_{x=1}^M \ \sum_{\substack{ 1 \leqslant i < j \leqslant N \\ 1 \leqslant k < l \leqslant N }} J'_{ijkl,x} \ \chi_{i,x} \chi_{j,x} \chi_{k,x+1} \chi_{l,x+1} 
	\eeqn
	where $\{J_{ijkl,x}\}$ and $\{J'_{ijkl,x}\}$ are independent zero-mean Gaussain random couplings with variances
	\beq
	\overline{J_{ijkl,x}^2} = \f{3! J_0^2}{N^3}, \qquad \qquad \overline{J^{'2}_{ijkl,x}} = \f{J_1^2}{N^3}.
	\eeq
	Each site of the lattice, labeled by $x = 1, \dots, M$, contains $N$ Majorana fermions with on-site SYK interactions. Nearest-neighbor SYK sites also interact via 2-2 fermion interactions, as shown in the second summation in the Hamiltonian of Eq. \eqref{HChain}. The Majorana fermions satisfy the anti-commutation relation
	\beq \label{Cliff}
	\{\chi_{i,x},\chi_{j,y}\}= \delta_{xy} \delta_{ij},
	\eeq
	and periodic boundary conditions: $\chi_{i,0} \equiv \chi_{i,M}$.  The model's local properties can be described in terms of an effective coupling constant
	\beq
	J = \sqrt{J_0^2 + J_1^2}.
	\eeq
	The SYK chain model is solvable in the large-$N$ limit. At strong coupling ($N \gg \beta J \gg 1$), the model retains the interesting properties of the original SYK model, such as maximal chaos, extensive zero-temperature entropy, and local criticality. In addition, due to its spatial locality, the model exhibits new features, such as energy diffusion and the butterfly effect.
	
	The model at strong coupling describes a strongly correlated diffusive metal. Also the relation $D= v_B^2/2 \pi T$ between its temperature independent diffusion constant $D$ and its butterfly speed $v_B$ is consistent with proposals of incoherent metals.  The SYK chain model, at strong coupling, provides a rare example of strongly interacting chaotic lattice models that are analytically tractable. As noted in Ref. \cite{Gu2017}, ``it provides an interesting platform for studying various properties of strongly correlated systems, such as thermalization, entanglement propagation, dissipative transport, etc.'' 
	
	Ref. \cite{Gu2017b} studied the spread of R\'enyi entropy in the SYK chain model and discussed its thermalization properties. However, the eigenstate thermalization hypothesis (ETH) for the model has not yet been directly investigated. In the following sections, I investigate the eigenstate thermalization of the SYK chain model.
	
	The diffusion constant $D$ of the model is a function of its couplings strengths:
\beq \label{D}
	D = \f{2 \pi J_1^2}{3 \sqrt{2} J \alpha_K}
\eeq
	where $\alpha_K \approx 2.852$ \cite{Gu2017}. The diffusion of energy in the system, indicated by the constant $D$, arises due to the dynamics of the collective mode of the time reparametrization field, which is identified as light degrees of freedom in \cite{Gu2017b}. This collective mode is the most important low-energy, long-wavelength mode in the system. Gu et al. \cite{Gu2017} find that single Majorana fermions do not propagate between sites; only collective modes made up of pairs of fermions exhibit spatial dynamics.
	
In study of off-diagonal matrix elements of observables in the next section I identify the Thouless energy, 
\beq \label{ET}
	E_T = \f{\hbar D}{L^2},
\eeq
where $L$ is the size of the system, and investigate whether the Thouless energy agrees with the expression in Eq. \eqref{D}.

\section{Computational Setup} \label{Setup}
I construct the SYK chain model with $M$ sites, each containing $N$ Majorana fermions. Thus, the total number of Majorana fermions in the system is $NM$.

Ref. \cite{Gu2017b} could explicitly evaluate R\'enyi entanglement entropy, and infer slow thermalization behavior, in the case of two-site SYK with $M=2$. They resorted to the weak inter-site coupling limit, defined as 
\beq \label{weak}
\f{J_1^2}{J^2} \ll \f{1}{\beta J} \ll 1,
\eeq
to obtain many of their results. 

Here, I study the two-site SYK model with $M=2$ in the weak inter-site coupling limit. Additionally, I explore the SYK chain model with $M=3$ sites in the weak coupling regime, and with $M=4$ sites in three distinct coupling regimes.

Throughout this work, I set $J = \sqrt{5}$. For $M = 2, 3$, I use:
\begin{equation} \label{Jcases0}
	J_0 = \sqrt{4.99}, \quad J_1 = 0.1,
\end{equation}
which satisfies the weak coupling condition in Eq.~\eqref{weak}.

For $M = 4$, I examine the following three parameter sets:
\begin{enumerate}[(a)] \label{Jcases}
	\item $J_0 = 2$, $J_1 = 1$ \hfill (Stronger intra-site coupling)
	\item $J_0 = J_1 = \sqrt{5/2}$ \hfill (Equal inter- and intra-site coupling)
	\item $J_0 = 1$, $J_1 = 2$ \hfill (Stronger inter-site coupling)
\end{enumerate}

Note that even in case (b), where inter- and intra-site couplings are equal, the model differs from the original SYK model, because the couplings in the SYK chain are not all-to-all. The SYK chain includes both on-site interactions (parameterized by $J_0$) and nearest-neighbor inter-site couplings (parameterized by $J_1$). Therefore, one should not expect this model to directly reproduce ETH results from the original SYK model.

To construct the matrix representation of the Hamiltonian, I first relabel the Majorana fermions by a single index:
\begin{equation}
	\chi_{i,x} \rightarrow \chi_\alpha,
\end{equation}
with
\begin{equation}
	\alpha = i + (x - 1) N,
\end{equation}
where $1 \leq i \leq N$ and $1 \leq x \leq M$, so that $\alpha$ ranges from $1$ to $NM$. The $\chi_\alpha$ obey the Clifford algebra:
\begin{equation}
\{ \chi_\alpha, \chi_\beta \} = \delta_{\alpha \beta}.
\end{equation}
Because Majorana fermions are Hermitian ($\chi_\alpha^\dagger = \chi_\alpha$), I define complex fermion operators as
\begin{align}
	c_\alpha &= \chi_{2\alpha} - i \chi_{2\alpha - 1}, \\
	c^\dagger_\alpha &= \chi_{2\alpha} + i \chi_{2\alpha - 1}, \quad \alpha = 1, \ldots, NM/2,
\end{align}
which satisfy the canonical fermionic anticommutation relations:
\begin{equation}
	\{ c_\alpha, c_\beta \} = \{ c^\dagger_\alpha, c^\dagger_\beta \} = 0, \qquad \{ c_\alpha, c^\dagger_\beta \} = \delta_{\alpha \beta}.
\end{equation}
I define a vacuum state $\ket{0}$ such that
\begin{equation}
	c_\alpha \ket{0} = 0,
\end{equation}
and construct the Hilbert space basis using
\begin{equation}
	(c_1^\dagger)^{n_1} \cdots (c_L^\dagger)^{n_L} \ket{0}, \qquad n_\alpha = 0, 1,
\end{equation}
where $L = NM/2$. This yields a $2^{NM/2}$-dimensional Hilbert space.

The matrix representation of the $\chi_\alpha$ operators can then be constructed recursively~\cite{Sarosi2018}:
\begin{align}
	\chi_\beta^{(K)} &= \chi_\beta^{(K-1)} \otimes 
	\begin{pmatrix}
		-1 & 0 \\
		0 & 1
	\end{pmatrix}, \quad \beta = 1, \ldots, 2K - 2, \\
	\chi_{2K - 1}^{(K)} &= I_{2^{K-1}} \otimes 
	\begin{pmatrix}
		0 & 1 \\
		1 & 0
	\end{pmatrix}, \\
	\chi_{2K}^{(K)} &= I_{2^{K-1}} \otimes 
	\begin{pmatrix}
		0 & -i \\
		i & 0
	\end{pmatrix},
\end{align}
where $I_d$ denotes the $d \times d$ identity matrix and $K$ runs from $1$ to $NM/2$. I initialize the recursion with
\begin{equation}
	\chi_1^{(1)} = 
	\begin{pmatrix}
		0 & -i \\
		i & 0
	\end{pmatrix} = Y, \quad
	\chi_2^{(1)} = 
	\begin{pmatrix}
		0 & 1 \\
		1 & 0
	\end{pmatrix} = X.
\end{equation}
This generates the full set of $2^{NM/2} \times 2^{NM/2}$ matrices $\{\chi_\alpha\}$.

Once the matrices $\{\chi_\alpha\}$ are constructed, one can recover the original indexing:
\begin{equation}
	i = \alpha \bmod N, \qquad x = \frac{\alpha - i}{N} + 1.
\end{equation}
This allows us to construct the Hamiltonian of the SYK chain model as given in Eq.~\eqref{HChain}.

\begin{figure*}
	\centering
	\includegraphics[width=.48\linewidth]{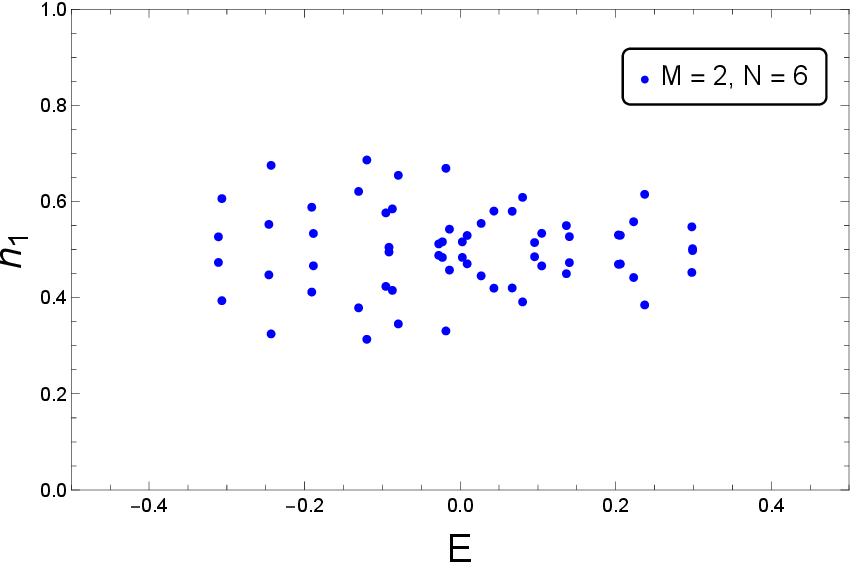}
	\includegraphics[width=.48\linewidth]{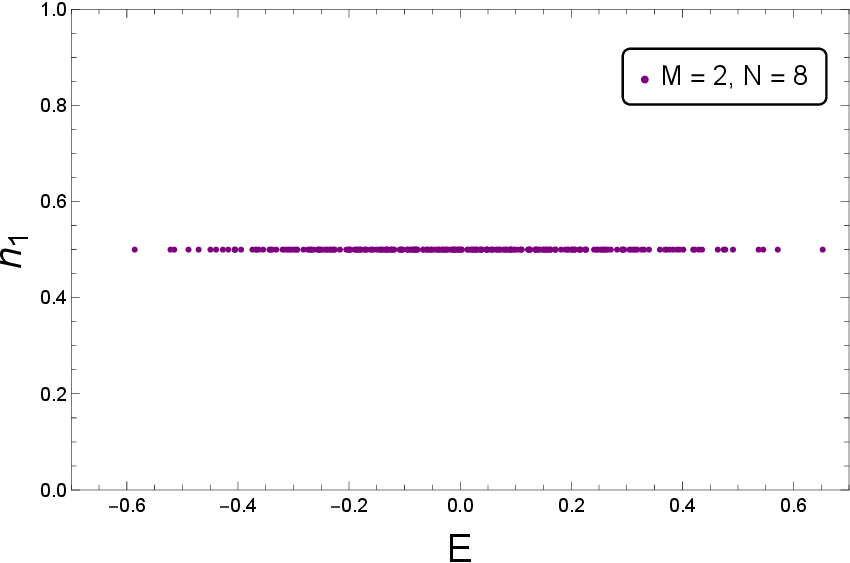}
	\includegraphics[width=.48\linewidth]{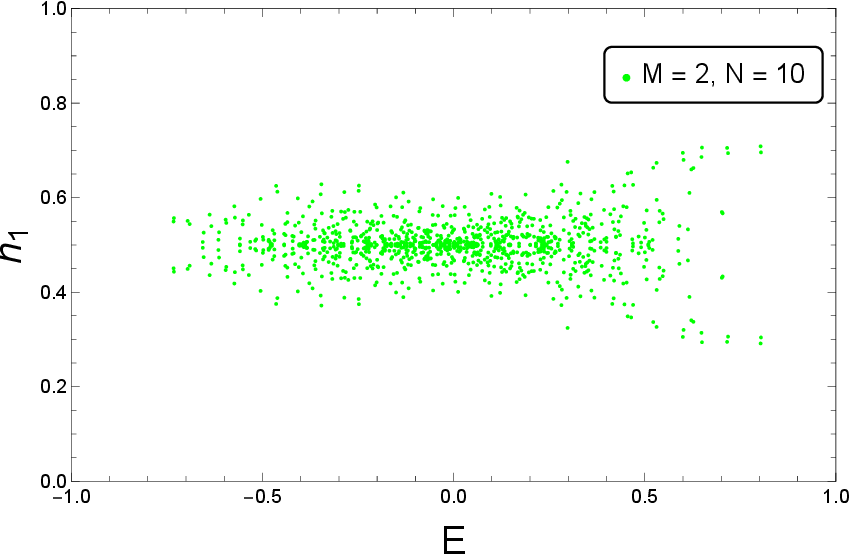}
	\includegraphics[width=.48\linewidth]{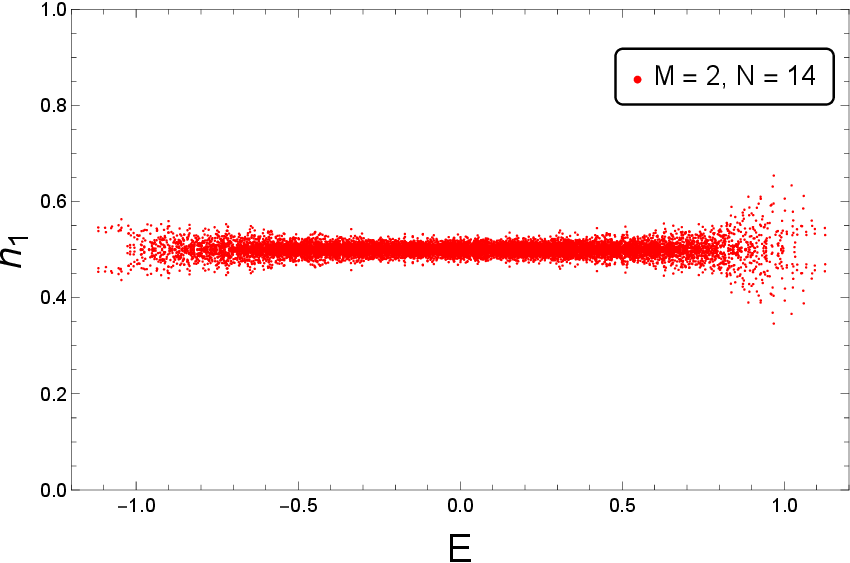}
	\caption{The absolute value of the on-diagonal elements of the particle number operator $n_1$ as a function of energy eigenvalues $E$ for single realization of two-site SYK model in the weak link limit. We consider $M N = 12, 20, 28$ Majorana fermions, with $M = 2$, and observe
		the diagonal terms fluctuating around 1/2.}
	\label{fig:diagonalenergym2n6-14}
\end{figure*}

\begin{figure*}
	\centering
	\includegraphics[width=0.48\linewidth]{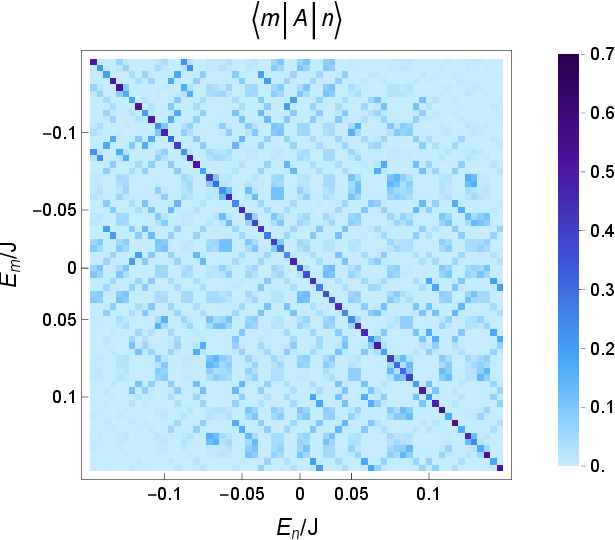}
	\includegraphics[width=0.48\linewidth]{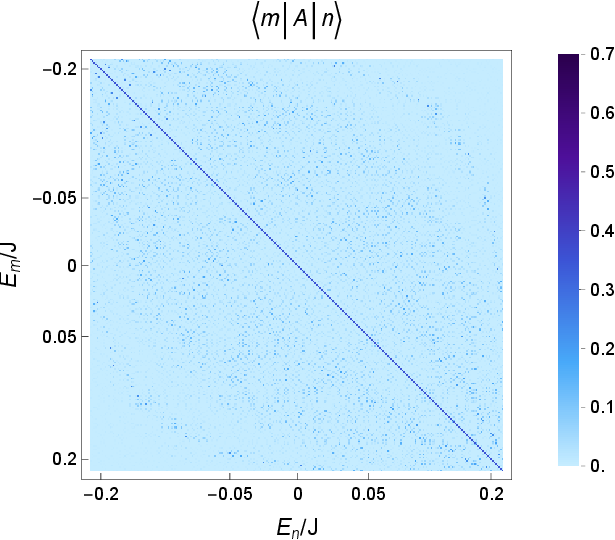}
	\includegraphics[width=0.48\linewidth]{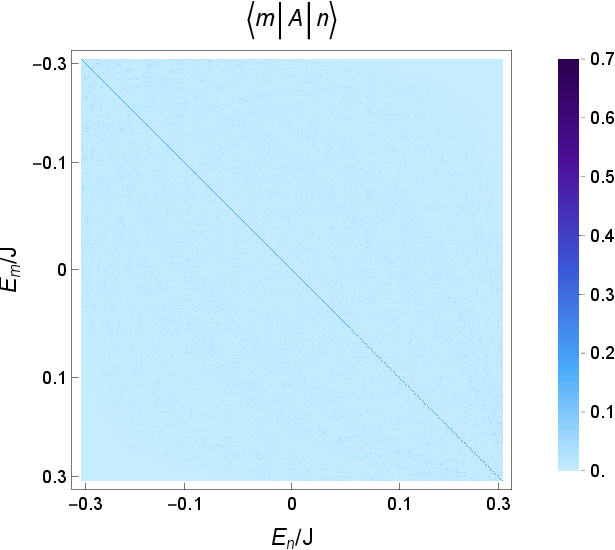}
	\includegraphics[width=0.48\linewidth]{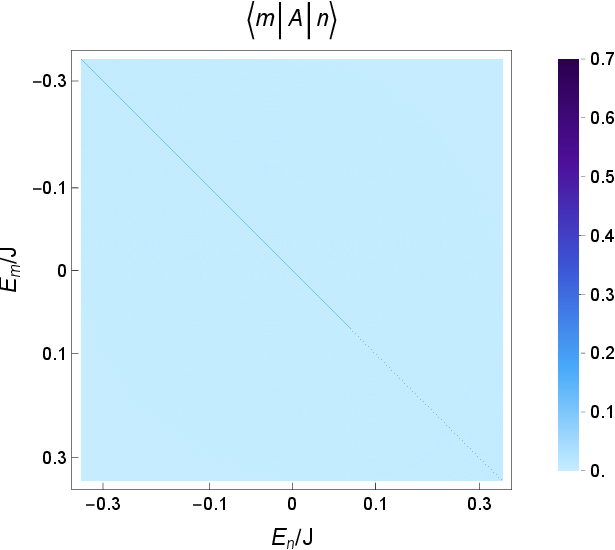}
	\caption{Matrix plots of $\lvert A_{mn}\rvert = \lvert \bra{m} A \ket{n} \rvert$ for the particle number operator $A =\hat{n}_1$ in the weak inter-site coupling regime, $J_0 = \sqrt{4.99}, J_1 = 0.1$, of the two-site SYK model, $M = 2$. I choose single realizations of the two-site SYK model with $M=2$ sites and $N=6, 8, 10, 14$ Majorana fermions on each site. The horizontal and vertical axes show the energy eigenvalues, $E_m/J$, $E_n/J$, corresponding to the matrix elements $\lvert \mathcal{O}_{mn}\rvert$. All three plots show agreement with ETH in its loose sense. The diagonal elements of the number operator matrices in the energy eigenbases fluctuate near the microcanonical ensemble average value $1/2$ while the off-diagonal ones are sub-exponentially small. This suggests that ETH is not satisfied in its strict sense but it is satisfied in its loose sense. \it{Top Left Panel:} $N = 6$, \it{Top Right Panel:} $N = 8$, \it{Bottom Left Panel:} $N = 10$, \it{Bottom Right Panel:} $N = 14$, the data for the latter are downsampled to 2048 $\times$ 2048 for the purpose of presentation.}	
	\label{fig:matrixplotm2}
\end{figure*}

\begin{figure*}[h!]
	\centering
	\includegraphics[width=.48\linewidth]{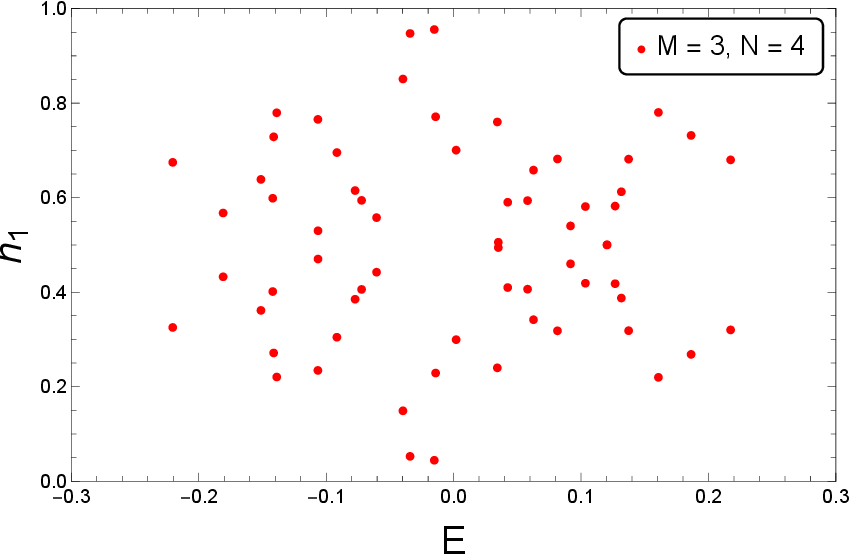}
	\includegraphics[width=.48\linewidth]{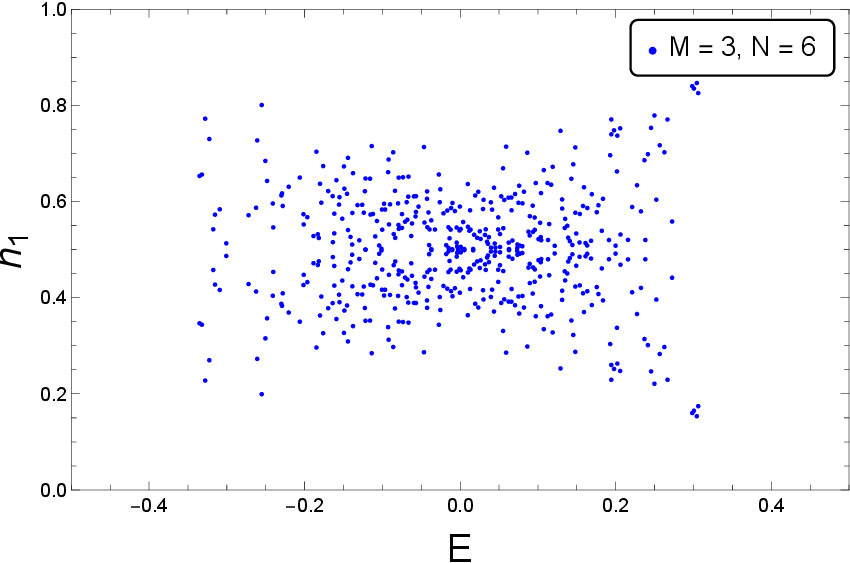}
	\includegraphics[width=.48\linewidth]{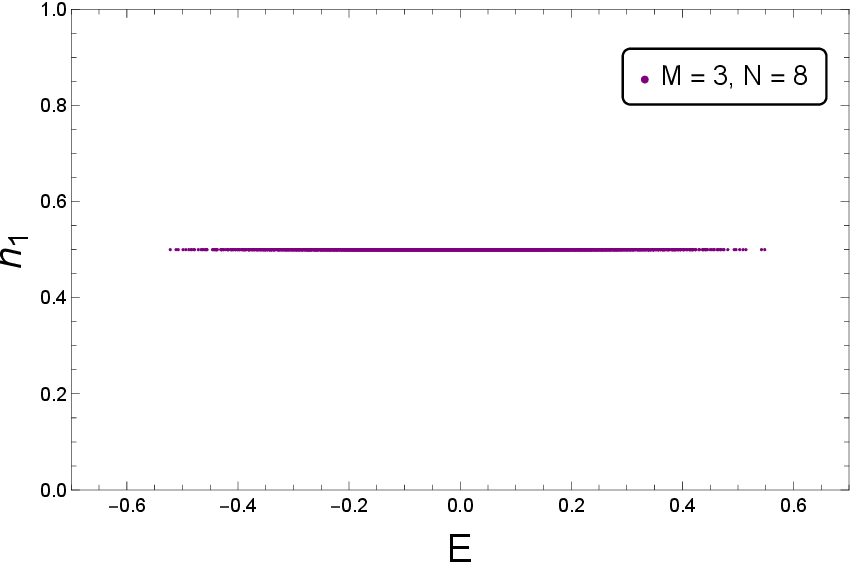}
	\includegraphics[width=.48\linewidth]{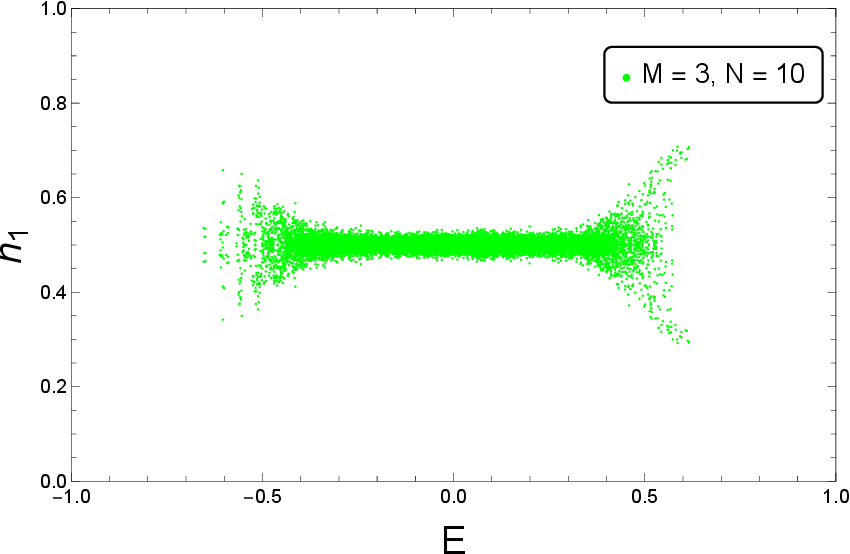}
	
	\caption{The absolute value of the on-diagonal elements of the particle number operator $n_1$ as a function of energy eigenvalues $E$ for single realizations of SYK chain model with three sites in the weak link regime. We considered $M N = 12, 18, 24, 30$ Majorana fermions, with $M = 3$, and observed
		the diagonal terms fluctuating around 1/2.}
	\label{fig:scatterm3}
\end{figure*}	

\begin{figure*}[h!]
	\centering
	\includegraphics[width=0.48\linewidth]{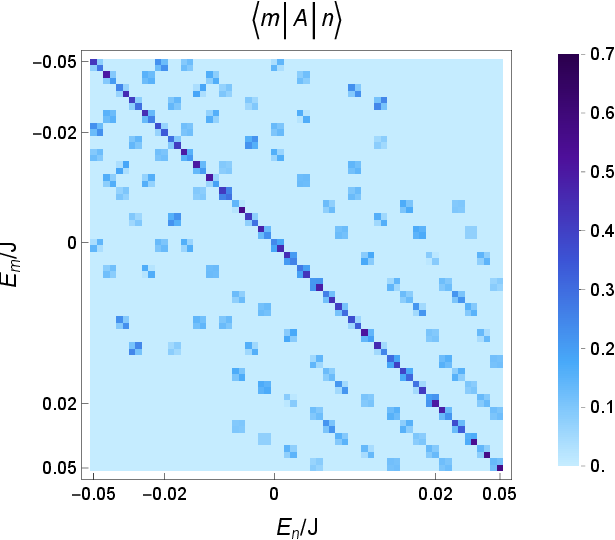}
	\includegraphics[width=0.48\linewidth]{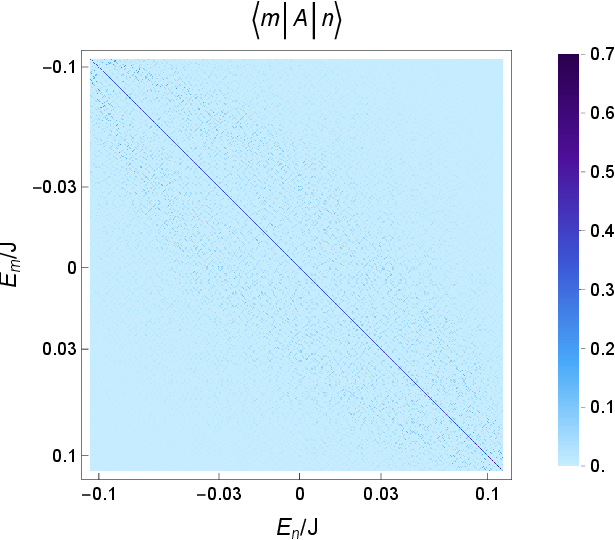}
	\includegraphics[width=0.48\linewidth]{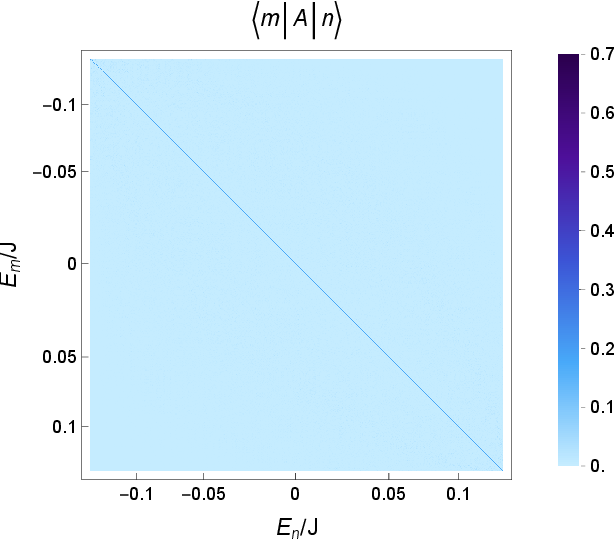}
	\includegraphics[width=0.48\linewidth]{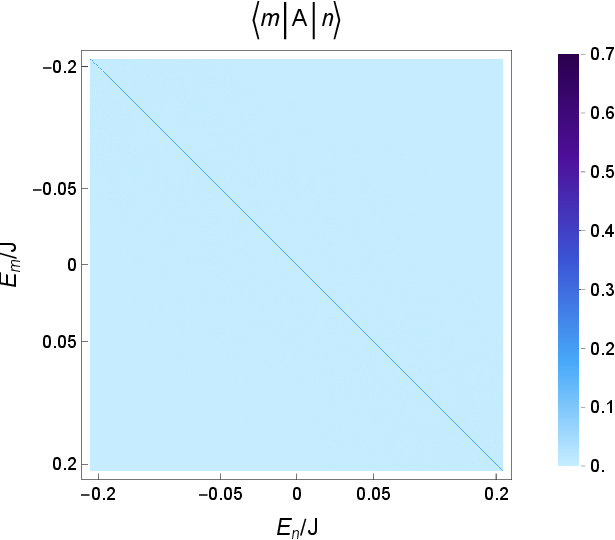}
	\caption{Matrix plots of $\lvert A_{mn}\rvert = \lvert \bra{m} A \ket{n} \rvert$ for the particle number operator $A =\hat{n}_1$ in the weak inter-site coupling regime $J_0 = \sqrt{4.99}, J_1 = 0.1$, of the SYK chain model with $M = 3$ sites. We choose single realizations of the model with $N= 4, 6, 8, 10$ Majorana fermions on each site. The horizontal and vertical axes show the energy eigenvalues, $E_m/J$, $E_n/J$, corresponding to the matrix elements $\lvert A_{mn}\rvert$. All three plots show agreement with ETH in its loose sense. The diagonal elements of the number operator matrices in the energy eigenbases fluctuate near the microcanonical ensemble average value $1/2$ while the off-diagonal ones are sub-exponentially small. This suggests that ETH is not satisfied in its strict sense but it is satisfied in its loose sense. \it{Top Left Panel:} $N = 4$, \it{Top Right Panel:} $N = 6$, \it{Bottom Left Panel:} $N = 8$, \it{Bottom Right Panel:} $N = 10$, the data for $N = 8, 10$ are downsampled to  1024 $\times$ 1024 for the purpose of presentation.}
	\label{fig:matrixplotm3}
\end{figure*}

What we did above was to map every two Majorana fermions (labeled by Latin indices) onto one complex fermion (labeled by Greek indices). In the language of complex fermions, we can say that there are \( M \) sites, and on each site, there can be \( N/2 \) complex fermions (where \( N \) is even here and throughout the paper). Therefore, the indices of complex fermions \( \alpha = 1, \cdots, N/2 \) belong to site \( x = 1 \); the indices \( \alpha = N/2 + 1, \cdots, N \) belong to site 2; the indices \( \alpha = N + 1, \cdots, 3N/2 \) belong to site 3; and so on. Finally, the indices of complex fermions \( \alpha = \frac{N (M-1)}{2} + 1, \cdots, \frac{NM}{2} \) belong to site \( x = M \). It is more convenient to express operators in terms of complex fermion operators, as I do below.\footnote{In another picture, one can map the complex fermions to qubits and regard fermion operators as gate operators that act on qubits.}

In the following sections, I examine ETH for two types of non-extensive operators: the particle number operator and the hopping operator. These operators act on one and two bodies, respectively. The particle number operator is constructed from the annihilation and creation operators as  
\begin{equation}
	\hat{n}_1 = c^\dagger_1 c_1,
\end{equation}  
where I use Greek indices to label complex fermion operators \( c_\alpha \), \( c^\dagger_\beta \). Physically, this corresponds to mapping the first two Majorana fermions on the first SYK site, \( \chi_{1,1} \) and \( \chi_{2,1} \), onto a single complex fermion. The number operator \( \hat{n}_1 \) then counts the occupation number of this fermion, which can be either 0 or 1. As defined, \( \hat{n}_1 \) is a local operator, influenced only by the first two Majorana fermions of the first site.

In contrast, the hopping operator can be completely non-local, acting between two distinct SYK sites. Using complex fermion indices again, I define the hopping operator between site 1 and site 3 as  
\begin{equation}
	\hat{h}_{13} = c^\dagger_{N+1} c_1 + c^\dagger_1 c_{N+1}.
\end{equation}  
Here, \( \alpha = 1 \) corresponds to a complex fermion on site \( x = 1 \), and \( \alpha = N + 1 \) corresponds to one on site \( x = 3 \). In terms of Majorana fermions, this operator involves the first two fermions of the first site, \( \chi_{1,1} \) and \( \chi_{2,1} \), and the first two of the third site, \( \chi_{2N+1,3} \) and \( \chi_{2N+2,3} \). Despite its non-locality, the hopping operator remains non-extensive, as it acts on only a few Majorana modes.

In the next sections, I present results for the particle number operator \( \hat{n}_1 \) and the hopping operator \( \hat{h}_{13} \) with different numbers of sites. 

The microcanonical average value of \( \hat{n}_1 \) is  
\begin{equation} \label{half}
	\langle \hat{n}_1 \rangle_{\text{micro}} = \frac{1}{2}.
\end{equation}  
According to ETH, as expressed in Eq.~\eqref{result11}, the diagonal matrix elements of \( \hat{n}_1 \) in the energy eigenbasis are expected to be approximately equal to this value.

In the case of off-diagonal matrix elements, the strict version of ETH, as given in Eq.~\eqref{ETH2}, predicts that all such elements should be of order \( e^{-S/2} \), where \( S \) is the thermodynamic entropy. In contrast, the weaker or approximate formulation of ETH, discussed in Sec.~\ref{OverviewETH}, requires only that the maximum absolute value of these off-diagonal elements be sufficiently small.

The entropy of a quantum system depends on its state: it is zero for pure states and maximal for the maximally mixed state, where the maximum value is \( \ln(\text{dim}(\mathcal{H})) \). For all other states, including thermal states, the entropy lies between these extremes. For the purpose of comparison in our numerical analysis, I use the mean value  
\begin{equation}
\bar{S} = \frac{1}{2} \ln(\text{dim}(\mathcal{H}))
\end{equation}   
to assess how small the off-diagonal matrix elements are relative to ETH expectations. For the two-site SYK and SYK chain models, the Hilbert space dimension is \( \text{dim}(\mathcal{H}) = 2^{NM/2} \), which gives the mean entropy as
\begin{equation}
\bar{S} = NM \ln 2 / 2.
\end{equation}

\section{Thermalization of Single Realizations of Two-site SYK and SYK Chain Models} \label{NumericalChecks-Single}
	
	The two-site SYK and SYK chain models are formulated as ensemble-averaged theories. However, the Eigenstate Thermalization Hypothesis (ETH) applies to individual quantum systems, not ensemble averages. Therefore, it is crucial to verify whether ETH holds for single realizations of these models before considering the ensemble-averaged versions.
	
	If single realizations of the disorder do not satisfy ETH, drawing any conclusions from the satisfaction of ETH in the disorder-averaged theories becomes problematic. This is because the proof of thermalization for generic observables within the ETH framework relies on the assumption that the system of interest is an individual system with a fixed Hamiltonian.
	
	In this section, I numerically examine the validity of the Eigenstate Thermalization Hypothesis for single realizations of the two-site SYK and SYK chain models. 
	
	The numerical results indicate that, for the operators examined in this study, ETH is \textit{approximately} satisfied in individual realizations of the disorder (see Section~\ref{OverviewETH} for the definition of \textit{approximate satisfaction of ETH}). Consequently, the analysis of ETH in the ensemble-averaged formulations of the two-site SYK and SYK chain models, as pursued in the subsequent section, remains well justified.

\subsection{Thermalization in Single Realizations of the Two-Site SYK Model}

In the weak coupling regime, introduced in Sec.~\ref{Jcases}, I examined single realizations of the two-site SYK model with $M = 2$ and $N = 6,\ 8,\ 10,\ 14$, using coupling strengths $J_0 = \sqrt{4.99}$ and $J_1 = 0.1$ for inter-site and intra-site interactions, respectively. I computed the absolute values of matrix elements of the particle number operator $\hat{n}_1$ in the energy eigenbasis, namely $|\langle m | \hat{n}_1 | n \rangle|$. The diagonal matrix elements were observed to fluctuate around the microcanonical average value of $1/2$ (as given in Eq.~\eqref{half}), while the off-diagonal elements remained relatively small.

Figure~\ref{fig:diagonalenergym2n6-14} displays the diagonal matrix elements as a function of energy eigenvalues. With the exception of $N = 8$, these elements fluctuate about $1/2$, and their deviation from this thermal value diminishes as $N$ increases.

For $N = 8$, the diagonal elements are exactly equal to $1/2$, a feature attributable to symmetry, as discussed in the context of the original SYK model (see Ref.~\cite{Hunter2018}).

Figure~\ref{fig:matrixplotm2} presents matrix plots of $|\langle m | \hat{n}_1 | n \rangle|$, clearly showing dark blue diagonal elements (corresponding to values near $0.5$) and light cyan off-diagonal elements (corresponding to values close to zero). In many cases, the off-diagonal elements vanish to machine precision, which is consistent with the block-diagonal structure of the Hamiltonian, similar to the original SYK model. Nevertheless, most non-zero off-diagonal elements are not as small as $e^{-\bar{S}/2}$, indicating a failure of strict ETH.

Despite this, both the mean and maximum values of the off-diagonal matrix elements decrease with increasing $N$. Table~\ref{maxmean} summarizes these values and compares them with the ETH-predicted value $e^{-\bar{S}/2} = e^{-NM \ln 2 / 4}$.

\begin{table}[htbp]
	\centering
	\caption{Maximum and mean values of the off-diagonal matrix elements of the particle number operator $A = \hat{n}_1$ in the energy eigenbasis for typical single realizations of the two-site SYK model ($M=2$), with different numbers of Majorana fermions $N$ per site.}
	\vspace{0.3cm}
	\begin{tabular}{|l|l|c|c|c|}
		\hline
		\textbf{$M$} & \textbf{$N$} & \textbf{$\max_{n \ne m} |A_{nm}|$} & \text{mean$_{n \ne m} |A_{nm}|$} & $e^{-\bar{S}/2}$\\ 
		\hline
		2 & 6             & 0.40    & 0.023 & 0.125 \\ 
		2 & 8             & 0.37    & 0.010 & 0.063 \\ 
		2 & 10            & 0.24    & 0.007 & 0.031 \\
		2 & 14            & 0.15    & 0.002 & 0.008 \\
		\hline
	\end{tabular}
	\label{maxmean}
\end{table}

As the data in Table~\ref{maxmean} illustrate, the strict formulation of ETH is not fulfilled, as $\max_{n \ne m} |A_{nm}| \gg e^{-\bar{S}/2}$ even for large $N$. However, the approximate form of ETH is satisfied: the off-diagonal matrix elements remain small, and their magnitude decreases with increasing $N$.

Moreover, the mean values of the off-diagonal elements approach or fall below the ETH expected value, mean$_{n \ne m}[\abs{A_{nm}}] \lesssim \exp(-\bar{S}/2)$, suggesting that ETH is satisfied in the ensemble-averaged framework.

\subsection{Thermalization in Single Realizations of the Three-Site SYK Chain Model} \label{SingleThreeSite}

For the three-site SYK chain ($M = 3$), I investigated single realizations with $N = 4,\ 6,\ 8,\ 10$ in the same weak coupling regime ($J_0 = \sqrt{4.99}, J_1 = 0.1$). As before, I computed $|\langle m | \hat{n}_1 | n \rangle|$ and found that the diagonal elements typically fluctuate around $1/2$ for $N = 4,\ 6,\ 10$. For $N = 8$, the elements are again exactly equal to $1/2$, reflecting the same symmetry mentioned earlier. These results are shown in Fig.~\ref{fig:scatterm3}.

Figure~\ref{fig:matrixplotm3} provides matrix plots of $|\langle m | \hat{n}_1 | n \rangle|$. As in the two-site case, the diagonal elements exhibit values near $0.5$, while most off-diagonal elements are close to zero or vanish to machine precision. This again reflects the block-diagonal structure of the Hamiltonian. However, the non-zero off-diagonal matrix elements generally exceed the ETH prediction $e^{-\bar{S}/2}$.

Table~\ref{maxmeanm3} presents the mean and maximum values of the off-diagonal elements for various values of $N$, together with the corresponding ETH prediction.

\begin{table}
	\centering
	\caption{Maximum and mean values of the off-diagonal matrix elements of the particle number operator $A = \hat{n}_1$ in the energy eigenbasis for typical single realizations of the three-site SYK chain ($M = 3$), with different numbers of Majorana fermions $N$ per site.}
	\vspace{0.3cm}
	\begin{tabular}{|l|l|c|c|c|}
		\hline
		\textbf{$M$} & \textbf{$N$} & \textbf{$\max_{n \ne m} |A_{nm}|$} & \text{mean$_{n \ne m} |A_{nm}|$} & $e^{-\bar{S}/2}$\\ 
		\hline
		3 & 4             & 0.37    & 0.016 & 0.125 \\
		3 & 6             & 0.34    & 0.007 & 0.044 \\ 
		3 & 8             & 0.30    & 0.002 & 0.016 \\ 
		3 & 10            & 0.22    & 0.001 & 0.006 \\
		\hline
	\end{tabular}
	\label{maxmeanm3}
\end{table}

The results in Table~\ref{maxmeanm3} confirm that strict ETH is not satisfied, as $\max_{n \ne m} |A_{nm}| \gg e^{-\bar{S}/2}$ for all cases. Nevertheless, the approximate form of ETH holds: the off-diagonal elements are small and decrease with $N$.

Additionally, the mean values of the off-diagonal matrix elements approach or fall below the ETH expected value, mean$_{n \ne m}[\abs{A_{nm}}] \lesssim \exp(-\bar{S}/2)$, indicating that ETH is satisfied for the ensemble-averaged SYK chain with three sites.

	\subsection{Thermalization of Single Realizations of the SYK Chain Model with Four Sites} \label{sec:SingleRealM4}
	
		\begin{figure*}
		\centering
		\includegraphics[width=.48\linewidth]{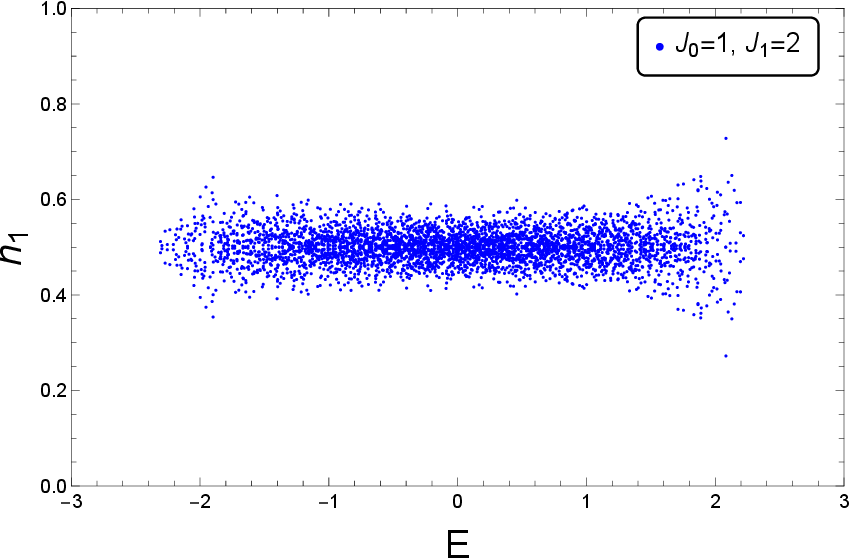}
		\includegraphics[width=.48\linewidth]{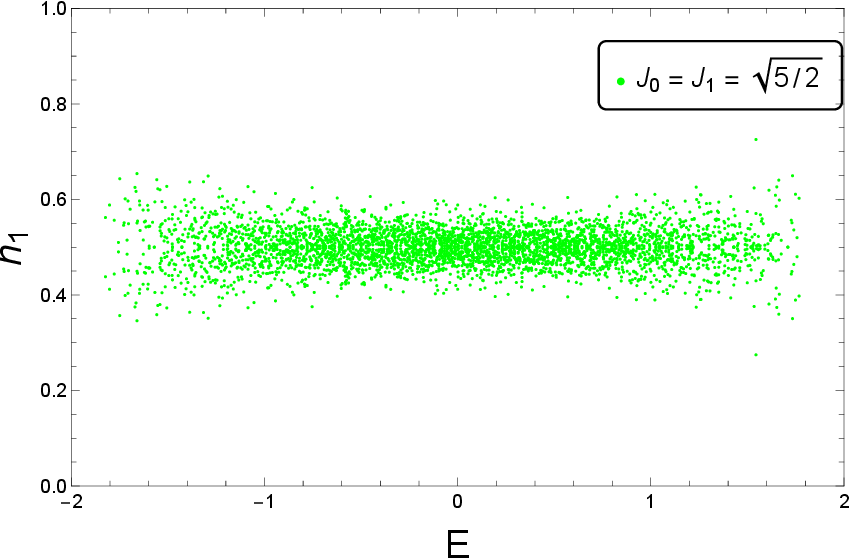}
		\includegraphics[width=.48\linewidth]{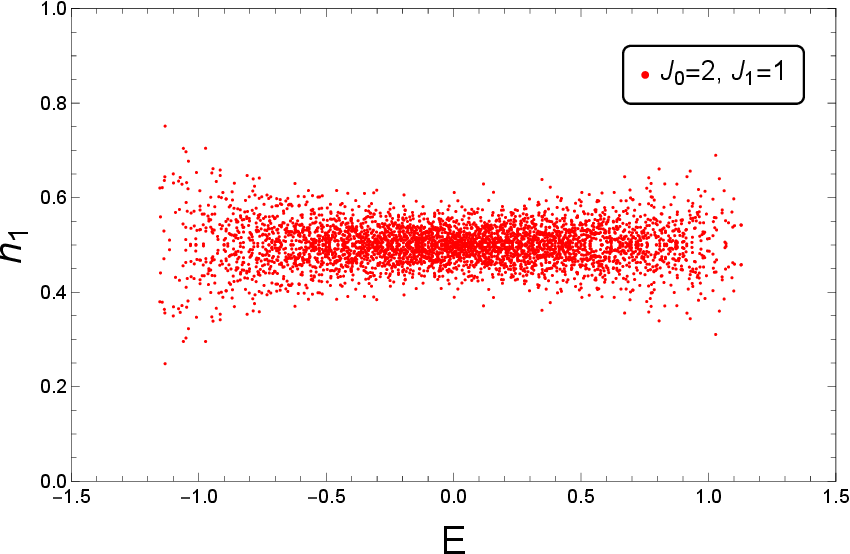}
		\caption{The absolute value of the on-diagonal elements of the particle number operator $n_1$ as a function of energy eigenvalues $E$ for single realization of SYK chain model with $M = 4$ sites and $N = 6$ fermions on each site. We consider three regimes of coupling strengths and observe that the diagonal terms fluctuate around the thermal value of $1/2$ in all cases.  \it{Top Left Panel:} $J_0= 1$, $J_1 = 2$, $J = \sqrt{5}$, \it{Top Right Panel:} $J_0= J_1 = \sqrt{5/2}$, $J = \sqrt{5}$, \it{Bottom Panel:} $J_0= 2$, $J_1 = 1$, $J = \sqrt{5}$. }
		\label{fig:diagonalenergym4}
	\end{figure*}

	\begin{figure*}[h!]
		\centering
		\includegraphics[width=0.48\linewidth]{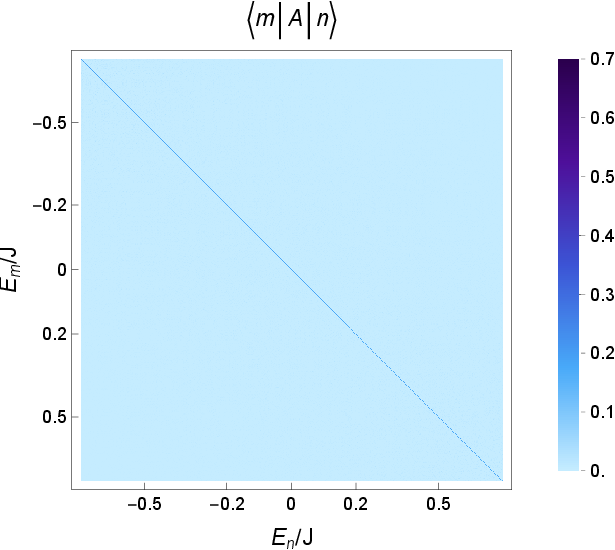}
		\includegraphics[width=0.48\linewidth]{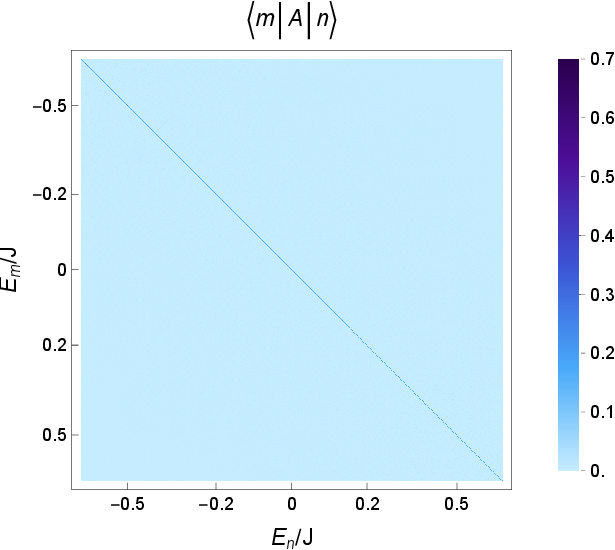}
		\includegraphics[width=0.48\linewidth]{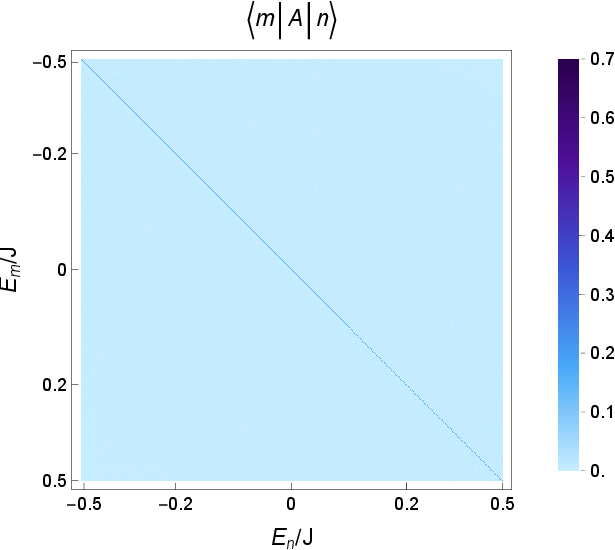}
		\caption{Matrix plots of $\lvert A_{mn}\rvert = \lvert \bra{m} A \ket{n} \rvert$ for the particle number operator $A =\hat{n}_1$ for three different values of coupling strengths $J_0,J_1$. We choose a single realization of the SYK chain model with $M=4$ sites and $N=6$ Majorana fermions on each site. The horizontal and vertical axes show the energy eigenvalues, $E_m/J$, $E_n/J$, corresponding to the matrix elements $\lvert A_{mn}\rvert$. All three plots show agreement with ETH in its loose sense. The diagonal elements of the number operator matrices in the energy eigenbases fluctuate near the microcanonical ensemble average value $1/2$ while the off-diagonal ones are zero or small. The non-zero off-diagonal matrix elements fluctuate around the expected exponentially small value of ETH by about two orders of magnitude. The maximum absolute value of the off-diagonal matrix elements are about $0.16$. This suggests that ETH is not satisfied in its strict sense but it is satisfied in its loose sense. For the purpose of presentation, the data are downsampled to 1024 $\times$ 1024. \it{Top Left Panel:} $J_0= 1$, $J_1 = 2$, \it{Top Right Panel:} $J = \sqrt{5}$, $J_0= J_1 = \sqrt{5/2}$, $J = \sqrt{5}$, \it{Bottom Panel:} $J_0= 2$, $J_1 = 1$, $J = \sqrt{5}$.  }	
		\label{fig:matrixplotm4}
	\end{figure*}

Turning to the four-site SYK chain ($M = 4$), I studied single realizations with $N = 6$ Majorana fermions per site for the three types of couplings described in Sec.\ref{Jcases}: $J_0 = 2, J_1 = 1$; $J_0 = J_1 = \sqrt{5/2}$; and $J_0 = 1, J_1 = 2$. As in previous cases, I computed $|\langle m | \hat{n}_1 | n \rangle|$ and observed that the diagonal elements fluctuate around $1/2$, as shown in Fig.\ref{fig:diagonalenergym4}.

Figure~\ref{fig:matrixplotm4} presents matrix plots of $|\langle m | \hat{n}_1 | n \rangle|$. The pattern remains consistent with that found for two- and three-site chains: diagonal entries cluster around $0.5$, while off-diagonal entries are mostly small or vanish at machine precision. This reflects the block-diagonal nature of the Hamiltonian. However, the nonzero off-diagonal matrix elements generally exceed the ETH prediction $e^{-\bar{S}/2}$.

A quantitative summary of the mean and maximum off-diagonal matrix elements of $A = \hat{n}_1$ is provided in Table~\ref{maxmeanm4n1}, alongside the corresponding ETH estimates. These results reveal that strict ETH is violated, with $\max_{n \ne m} |A_{nm}| \gg e^{-\bar{S}/2}$ in all coupling scenarios. Nonetheless, an approximate version of ETH remains valid: the off-diagonal elements are small compared to unity.

\begin{table}[htbp]
	\centering
	\caption{
		Maximum and mean values of the off-diagonal matrix elements of the particle number operator \( A = \hat{n}_1 \) in the energy eigenbasis for representative single realizations of the SYK chain model with \( M = 4 \) sites and \( N = 6 \) Majorana fermions per site at three different coupling strengths.
	}
	\vspace{0.3cm}
	\begin{tabular}{|c|c|c|c|c|}
		\hline
		\textbf{$J_0$} & \textbf{$J_1$} & \textbf{$\max_{n \ne m} \abs{A_{nm}}$} & \text{mean$_{n \ne m} \abs{A_{nm}}$} & $e^{-\bar{S}/2}$ \\ 
		\hline
		2 & 1 & 0.16 & 0.002 & 0.016 \\ 
		$\sqrt{5/2}$ & $\sqrt{5/2}$ & 0.17 & 0.002 & 0.016 \\ 
		1 & 2 & 0.20 & 0.002 & 0.016 \\
		\hline
	\end{tabular}
	\label{maxmeanm4n1}
\end{table}

Moreover, the mean off-diagonal matrix elements tend to approach or fall below the ETH estimate, $\text{mean}_{n \ne m}[\abs{A_{nm}}] \lesssim e^{-\bar{S}/2}$. This supports the view that ETH holds in the ensemble-averaged sense for the four-site SYK chain. It is worth noting that while the maximum off-diagonal element varies with the coupling strength and realization, the mean value remains largely unaffected by the choice among the three coupling types considered.

Turning next to the hopping operator $\hat{h}_{13}$, a similar analysis yields consistent findings. The diagonal elements of $\hat{h}_{13}$ vanish exactly by symmetry:

\beq \label{hopping-diagonal} \langle n | \hat{h}_{13} | n \rangle = 0. \eeq

This follows directly from the fermionic structure of the model and the Hermiticity of the operator. Consequently, the microcanonical average also vanishes:

\beq \label{hopping-micro} 
\langle \hat{h}_{13} \rangle_\text{micro} = 0. 
\eeq

Numerically, I confirmed the strict satisfaction of Eq.~\eqref{hopping-diagonal} for all three coupling cases. Given that the microcanonical average is also zero, the first ETH condition--the agreement between diagonal matrix elements and the microcanonical average--is exactly satisfied for $\hat{h}_{13}$.

For the off-diagonal elements, Table~\ref{maxmeanm4h13} compiles the mean and maximum values of $|\mathcal{O}{nm}|$ for $\mathcal{O} = \hat{h}_{13}$ across the different coupling scenarios, together with the ETH estimates. As before, strict ETH is not satisfied, since $\max_{n \ne m} |\mathcal{O}_{nm}| \gg e^{-\bar{S}/2}$ in every case. Nevertheless, the off-diagonal elements remain small overall, consistent with an approximate ETH.

\begin{table}[htbp]
	\centering
	\caption{
		Maximum and mean values of the off-diagonal matrix elements of the hopping operator $\mathcal{O} = \hat{h}_{13}$ in the energy eigenbasis for typical single realizations of the SYK chain model with $M=4$ sites and $N=6$ Majorana fermions on each site for three different coupling strengths. 
	}
	\vspace{0.3cm}
	\begin{tabular}{|c|c|c|c|c|}
		\hline
		\textbf{$J_0$} & \textbf{$J_1$} & \textbf{$\max_{n \ne m} \abs{\mathcal{O}_{nm}}$} & \text{mean$_{n \ne m} \abs{\mathcal{O}_{nm}}$} & $e^{-\bar{S}/2}$ \\ 
		\hline
		2 & 1 & 0.20 & 0.003 & 0.016 \\ 
		$\sqrt{5/2}$ & $\sqrt{5/2}$ & 0.18 & 0.003 & 0.016 \\ 
		1 & 2 & 0.23 & 0.003 & 0.016 \\
		\hline
	\end{tabular}
	\label{maxmeanm4h13}
\end{table}

Again, the mean off-diagonal values approach or fall below $e^{-\bar{S}/2}$, reinforcing the conclusion that ETH holds in the ensemble-averaged four-site SYK chain. It should be emphasized that although the maximum off-diagonal element exhibits sensitivity to the coupling and realization, the mean value remains remarkably stable across the different coupling cases.
	
\section{Thermalization of Ensemble-Averaged Theories of the Two-Site SYK and SYK Chain Models} \label{NumericalChecks-Ensemble}

In the preceding section, I examined the ETH conditions for individual realizations of the two-site SYK and SYK chain models and found them to be approximately satisfied. These results suggest that single realizations of these models thermalize rapidly with respect to the observables considered, with subexponentially small fluctuations of their expectation values around the corresponding thermal values.

I now turn to the ensemble-averaged forms of the two-site SYK and SYK chain models, which exhibit additional interesting properties. A natural question arises: what can be said about the thermalization of these ensemble-averaged theories? Unfortunately, the ETH is formulated only for individual quantum systems, and no developed extension currently exists that can be directly applied to ensemble-averaged models.

Nevertheless, it is common practice to perform ETH-inspired analyses in such settings. In what follows, I adopt the procedure employed in Refs.~\cite{Sonner2017, Hunter2018} for studying the original SYK, supersymmetric SYK, and complex SYK models.

The analysis of diagonal matrix elements proceeds as follows. First, I partition the energy spectrum into small intervals. For each realization of the disorder, I compute the diagonal matrix elements of a chosen operator $\mathcal{O}$ in the energy eigenbasis, $\mathcal{O}_{nn} = \bra{n} \mathcal{O} \ket{n}$. I then average the $\mathcal{O}_{nn}$ corresponding to eigenstates whose energies $E_n$ lie within the same interval. Finally, I average over disorder realizations and report the resulting quantity as the disorder-averaged diagonal matrix element of $\mathcal{O}$ as a function of energy. I compare this result to the expectation from ETH.

For off-diagonal matrix elements, ETH predicts that their typical magnitudes should scale as $e^{-S/2}$:
\begin{equation} \label{gO}
	g_{\mathcal{O}} (\bar{E}, \omega) := \bra{m} \mathcal{O} \ket{n} \sim e^{-S/2}.
\end{equation}
Here, $g_{\mathcal{O}}$ denotes the off-diagonal matrix elements of $\mathcal{O}$, classified as a function of the average energy $\bar{E} = (E_m + E_n)/2$ and the energy difference $\omega = E_m - E_n$.

The analysis of off-diagonal matrix elements proceeds in a similar manner. First, I partition the $\bar{E}$ and $\omega$ domains into small intervals of widths $\delta \bar{E}$ and $\delta \omega$, respectively. For each realization of the disorder, I compute the off-diagonal matrix elements $g_{\mathcal{O}}(\bar{E}, \omega)$. I then group all matrix elements $g_{\mathcal{O}}(\bar{E}, \omega)$ whose corresponding $(\bar{E}, \omega)$ values fall within the same intervals $(\bar{E}, \bar{E} + \delta \bar{E})$ and $(\omega, \omega + \delta \omega)$, and compute their average. Finally, I average the results over disorder realizations and report the resulting quantity $\bar{g}_{\mathcal{O}}(\bar{E}, \omega)$ as the disorder-averaged off-diagonal matrix elements of $\mathcal{O}$. I then compare these results to the ETH expectation.

In equation \eqref{gO}, I expressed the off-diagonal matrix elements of the operator $\mathcal{O}$ as a function of the average energy $\bar{E}$ and the energy difference $\omega$. This parametrization enables a systematic study of the chaotic and random matrix theory (RMT)-like features of the off-diagonal elements. In RMT, at fixed $\bar{E}$, $g_{\mathcal{O}}(\bar{E}, \omega)$ is expected to be approximately constant as a function of $\omega$. Thus, to compare the behavior of the SYK chain model with that of a random matrix theory, I fix $\bar{E}$ and plot $g_{\mathcal{O}}(\bar{E}, \omega)$ as a function of $\omega$.

The portion of $g_{\mathcal{O}}(\bar{E}, \omega)$ that remains nearly constant with respect to $\omega$ exhibits RMT-like behavior, while deviations from constancy indicate non-RMT behavior. The crossover between these two regimes occurs at a characteristic energy scale denoted $E_T$. In local models with spatial structure, this energy scale is often associated with the diffusive Thouless energy. Since the SYK chain model possesses one-dimensional spatial structure and exhibits diffusion of collective modes \cite{Gu2017}, $E_T$ can be identified with the Thouless energy of the model. A similar identification has even been proposed for the complex, original, and supersymmetric SYK models, despite their zero-dimensional nature \cite{Sonner2017, Hunter2018}.

The SYK chain model is a one-dimensional system, and its diffusion constant has been theoretically derived, as shown in Eq. \eqref{D}. To take the analysis further, I investigate whether the Thouless energy, which is proportional to the diffusion constant (see Eq. \eqref{ET}), follows the dependence of the diffusion constant on the coupling strengths, as expressed in Eq. \eqref{D}. Specifically, I check if the relationship
\begin{equation} \label{ET2} 
	E_T \propto D,
\end{equation}
holds, with the diffusion constant given by
\begin{equation} \label{D2} 
	D = \frac{2 \pi J_1^2}{3 \sqrt{2} J \alpha_K}.
\end{equation}

\subsection{Thermalization of the Ensemble-Averaged Two-Site SYK Model}

For $N = 6$, $10$, and $14$ the ETH conditions were examined in the ensemble-averaged theory of the two-site SYK model ($M = 2$) at the weak link regime with $J_0 = \sqrt{4.99}$ and $J_1 = 0.1$. The procedure outlined previously was applied to the number operator $\hat{n}_1$ and the hopping operator $\hat{h}_{12}$, where $\hat{h}_{12}$ acts on the first and second sites in a similar fashion to $\hat{h}_{13}$, which acts on the first and third sites. Averages were taken over a sufficient number of disorder realizations to ensure that the resulting plots were smooth, with further increases in the number of realizations yielding no improvement in plot quality. Specifically, $40,000$ realizations were used for $N = 6$, $1,000$ for $N = 10$, and $25$ for $N = 14$.

\begin{figure} \centering \includegraphics[width=.9\linewidth]{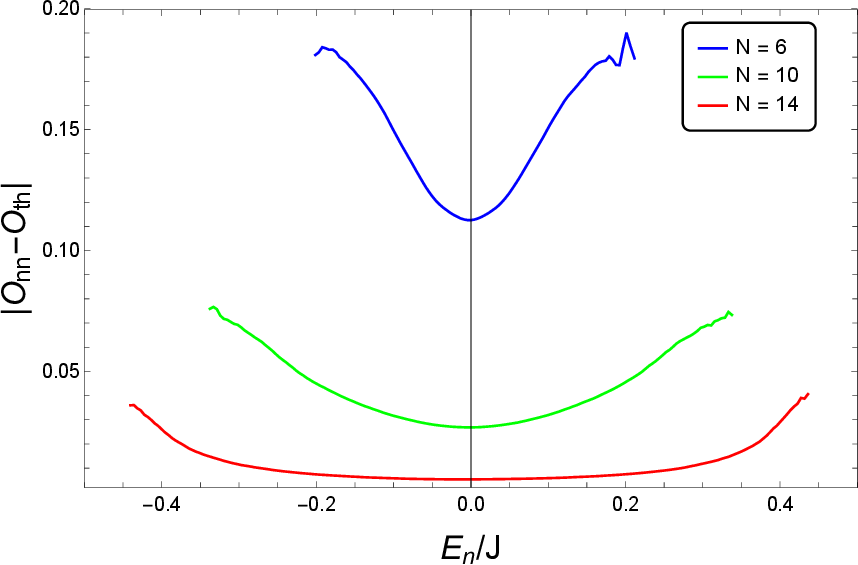} \caption{Absolute difference between the diagonal elements $\mathcal{O}_{nn}$ of the particle number operator, $\mathcal{O} = \hat{n}_1$, and its thermal value, $\mathcal{O}_{\mathrm{th}} = \mathrm{Tr}[\rho_{\mathrm{mc}} \mathcal{O}]/Z = 1/2$, plotted as a function of energy eigenvalues $E_n/J$ for the ensemble-averaged two-site SYK model with $N = 6$, $10$, and $14$ Majorana fermions per site at the weak-link regime with $J_0 = \sqrt{4.99}$ and $J_1 = 0.1$. Ensemble averages were taken over $40000$, $1000$, and $25$ disorder realizations, respectively. The smoothness of the curves and the small magnitude of the deviations demonstrate complete agreement with ETH for all three values of $N$.} \label{fig:diagnumvsenergym2} \end{figure}

The results for the diagonal matrix elements of the number operator $\mathcal{O} = \hat{n}_1$ are shown in Fig. \ref{fig:diagnumvsenergym2}. To generate the plot, the absolute difference between the diagonal matrix elements of the number operator in the energy eigenbasis and its microcanonical average value, $\frac{1}{2}$, was calculated for each realization, i.e., $\lvert \mathcal{O}_{nn} - \frac{1}{2} \rvert$. Averages were then computed over all disorder realizations using the procedure described earlier.

As observed, the ETH condition, as stated in Eq. \eqref{ETH1} or Eq. \eqref{result11}, is satisfied for the diagonal elements of the number operator. The deviations from the microcanonical average value are small, and they decrease further as $N$ increases. Therefore, ETH is strictly satisfied for the diagonal matrix elements of the number operator $\hat{n}_1$.

An interesting feature appears in Fig. \ref{fig:diagnumvsenergym2}. The deviation from the microcanonical average value is smaller at smaller absolute energies and becomes larger at larger absolute energies. This feature is also observed in the original SYK, supersymmetric SYK, and complex SYK models \cite{Sonner2017,Hunter2018}.

For the hopping operator $\hat{h}_{12}$, the diagonal matrix elements are strictly zero due to symmetry, just like $\hat{h}_{13}$, as discussed in Sec. \ref{sec:SingleRealM4}. Thus, their disorder averages are also zero and in strict agreement with their microcanonical value of zero.

\begin{figure} \centering \includegraphics[width=\linewidth]{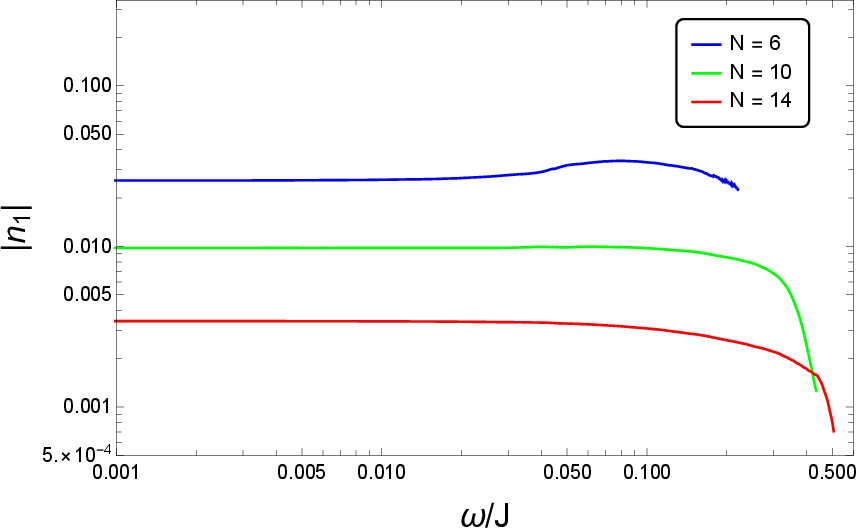} \caption{Absolute value of the off-diagonal elements $\mathcal{O}_{mn}$ of the particle number operator, $\mathcal{O} = \hat{n}_1$, plotted as a function of the energy difference $\omega/J = (E_m - E_n)/J$ at fixed average eigenenergy $(E_m + E_n)/2$ for $N = 6$, $10$, and $14$ Majorana fermions per site in the ensemble-averaged two-site SYK model at the weak-link regime with $J_0 = \sqrt{4.99}$ and $J_1 = 0.1$. Ensemble averages were performed over $40000$, $1000$, and $25$ disorder realizations, respectively. The exponentially small values of the off-diagonal elements demonstrate strict agreement with ETH for all three values of $N$.} \label{fig:offdiagonalnumopvsdem2} \end{figure}

\begin{figure} \centering \includegraphics[width=\linewidth]{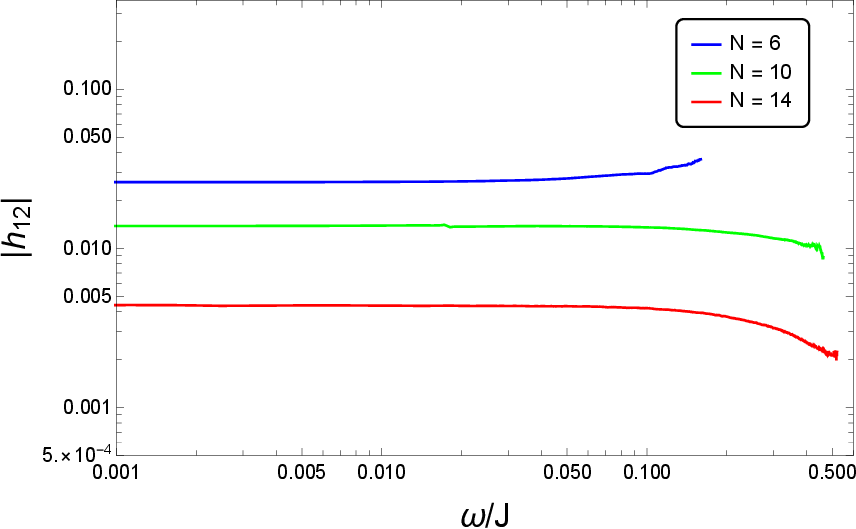} \caption{Absolute value of the off-diagonal elements $\mathcal{O}_{mn}$ of the hopping operator, $\mathcal{O} = \hat{h}_{12}$, plotted as a function of the energy difference $\omega/J = (E_m - E_n)/J$ at fixed average eigenenergy $(E_m + E_n)/2$ for $N = 6$, $10$, and $14$ Majorana fermions per site in the ensemble-averaged two-site SYK model at the weak-link regime with $J_0 = \sqrt{4.99}$ and $J_1 = 0.1$. Ensemble averages were performed over $40000$, $1000$, and $25$ disorder realizations, respectively. The exponentially small values of the off-diagonal elements demonstrate strict agreement with ETH for all three values of $N$.} \label{fig:offdiagonalhopopvsdem2} \end{figure}

For the off-diagonal matrix elements, however, there is no symmetry, and the values must be obtained numerically. The ensemble-averaged absolute values of the off-diagonal matrix elements of the number operator $\hat{n}_1$ and the hopping operator $\hat{h}_{12}$ are numerically computed and plotted in Figs. \ref{fig:offdiagonalnumopvsdem2} and \ref{fig:offdiagonalhopopvsdem2}.

As observed, the ETH expectations are satisfied for the off-diagonal matrix elements of both $\hat{n}_1$ and $\hat{h}_{12}$. The ensemble-averaged absolute values of the off-diagonal matrix elements are smaller than $e^{-\bar{S}/2}$ (see Table~\ref{maxmean} for the values of $e^{-\bar{S}/2}$). Thus, ETH is strictly satisfied for both operators in the ensemble-averaged two-site SYK model.

Finally, the Thouless energy $E_T$ can be identified in Figs.~\ref{fig:offdiagonalnumopvsdem2} and \ref{fig:offdiagonalhopopvsdem2} as the energy scale where the behavior transitions from being independent of $\omega$ (Random Matrix Theory (RMT) behavior) to being dependent on $\omega$ (non-RMT behavior). These results suggest that, for most of the energy domain, the two-site SYK model exhibits RMT-like behavior.

\subsection{Thermalization in the Ensemble-Averaged Three-Site SYK Chain}

I investigate thermalization in the ensemble-averaged SYK chain model with \( M = 3 \) sites and \( N = 8 \) Majorana fermions per site, averaging over 1000 disorder realizations. As established in Sec.~\ref{SingleThreeSite}, the diagonal matrix elements of the number operator \( \hat{n}_1 \) are identically equal to the microcanonical value \( 1/2 \) for \( N = 8 \), and the diagonal elements of the hopping operator \( \hat{h}_{13} \) are identically zero. Therefore, in this section, I focus exclusively on the off-diagonal matrix elements of \( \hat{n}_1 \) and \( \hat{h}_{13} \) in the ensemble-averaged three-site SYK chain. As in the analysis of single realizations, I consider the weak-link regime, with coupling constants \( J_0 = \sqrt{4.99} \) and \( J_1 = 0.1 \).

\begin{figure}
	\centering
	\includegraphics[width=\linewidth]{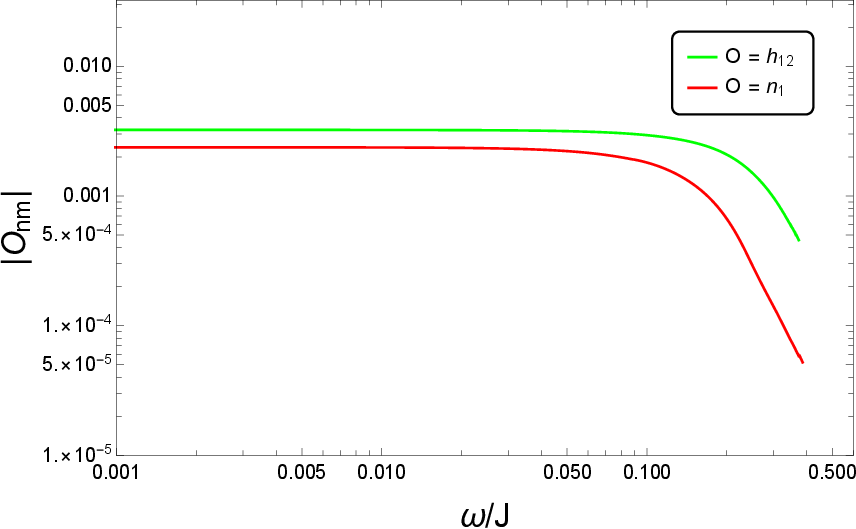}
	\caption{The ensemble-averaged absolute value of the off-diagonal matrix elements of the number operator \( \hat{n}_1 \) and the hopping operator \( \hat{h}_{13} \) as a function of energy difference \( \omega/J \) for the SYK chain with $M = 3$ sites and \( N = 8 \) Majorana fermions per site at the weak-link regime with $J_0 = \sqrt{4.99}$ and $J_1 = 0.1$. The average is taken over 1000 disorder realizations. The moving average over neighboring matrix elements is performed at fixed average eigenenergy. The results demonstrate strict agreement with the ETH predictions for both of the operators.}
	\label{fig:offdiagonalopvsdem3}
\end{figure}

The ensemble-averaged absolute values of the off-diagonal matrix elements of \( \hat{n}_1 \) and \( \hat{h}_{13} \) are computed numerically following the procedure outlined earlier and are presented in Fig.~\ref{fig:offdiagonalopvsdem3}.

As shown in Fig.~\ref{fig:offdiagonalopvsdem3}, the ensemble-averaged off-diagonal elements of both \( \hat{n}_1 \) and \( \hat{h}_{13} \) satisfy the ETH expectations. In particular, their magnitudes are consistently smaller than \( e^{-\bar{S}/2} \) (see Table~\ref{maxmeanm3} for numerical value of \( e^{-\bar{S}/2} \) for $M = 3$ and $N = 8$). Thus, the ETH holds strictly for both operators in the ensemble-averaged three-site SYK chain.

Moreover, Fig.~\ref{fig:offdiagonalopvsdem3} enables an estimation of the Thouless energy \( E_T \), identifiable as the scale where the behavior of the off-diagonal elements transitions from a flat (RMT-like) dependence to a decay with increasing \( \omega \). Interestingly, the location of this crossover appears to be largely independent of the choice of operator, reinforcing the validity of the theoretical analysis of Thouless energy in the SYK chain model given in Ref. \cite{Gu2017}.

\subsection{Thermalization in the Ensemble-Averaged Four-Site SYK Chain}

The ETH conditions are examined in the ensemble-averaged SYK chain with $M = 4$ sites and $N = 6$ Majorana fermions per site, across all three coupling regimes defined in Sec.~\ref{Jcases}, namely $J_0 = 2$, $J_1 = 1$; $J_0 = J_1 = \sqrt{5/2}$; and $J_0 = 1$, $J_1 = 2$. Following the procedure outlined earlier, the number operator $\hat{n}_1$ and the hopping operator $\hat{h}_{13}$ are analyzed, with averages taken over $1000$ disorder realizations.

\begin{figure}
	\centering
	\includegraphics[width=\linewidth]{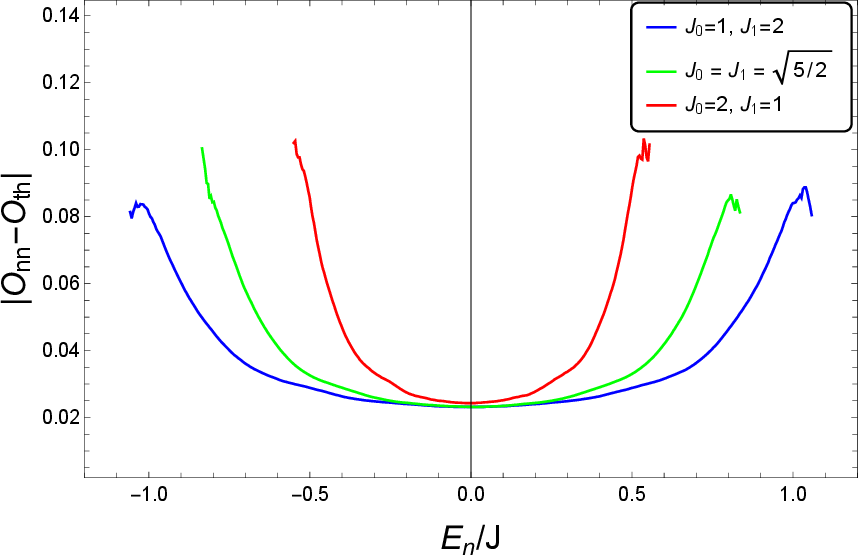}
	\caption{The absolute difference between the diagonal elements $\mathcal{O}_{nn}$ of the particle number operator, $\mathcal{O} = \hat{n}_1$, and its thermal value $\mathcal{O}_{\mathrm{th}} = \textrm{Tr} [\rho_{\mathrm{mc}} \mathcal{O}]/Z = 1/2$, plotted as a function of energy eigenvalues $E_n/J$ for three different coupling strengths $J_0$ and $J_1$. The ensemble average is taken over 1000 disorder realizations of the SYK chain model with $M=4$ sites and $N=6$ Majorana fermions per site. The curves exhibit a visible dependence on the ratio $J_1/J_0$, yet the smoothness of the curves and the smallness of the difference between the diagonal elements and the thermal value indicate complete agreement with ETH for all three pairs of coupling strengths.}
	\label{fig:diagnumvsenergyn4m6real1000}
	\end{figure}

The results for the diagonal matrix elements of the number operator, $\mathcal{O} = \hat{n}_1$, are shown in Fig.~\ref{fig:diagnumvsenergyn4m6real1000}.

As illustrated in Fig.~\ref{fig:diagnumvsenergyn4m6real1000}, the ETH expectation is satisfied for the diagonal elements of the number operator. The deviations from the microcanonical average value are small, and Eq.~\eqref{result11} is consequently satisfied. Thus, ETH holds for the diagonal matrix elements of $\hat{n}_1$ across all three coupling regimes.

Several notable features emerge from Fig.~\ref{fig:diagnumvsenergyn4m6real1000}. For all coupling cases, the deviation from the microcanonical value is smaller at lower absolute energies and grows larger at higher absolute energies. This behavior has also been observed in the original SYK model, the supersymmetric SYK model, and the complex SYK model~\cite{Sonner2017,Hunter2018}. The rate of growth of these deviations is faster in the case of stronger intra-site couplings ($J_0 = 2$, $J_1 = 1$) and becomes progressively slower as the inter-site couplings become stronger ($J_0 = 1$, $J_1 = 2$). This difference reflects the distinct roles played by intra-site and inter-site coupling terms in the SYK chain Hamiltonian, Eq.~\eqref{HChain}. However, since the deviations remain within the $\mathcal{O}(1/N)$ domain permitted by ETH, they do not affect the overall validity of ETH in this system.

\begin{figure}
	\centering
	\includegraphics[width=\linewidth]{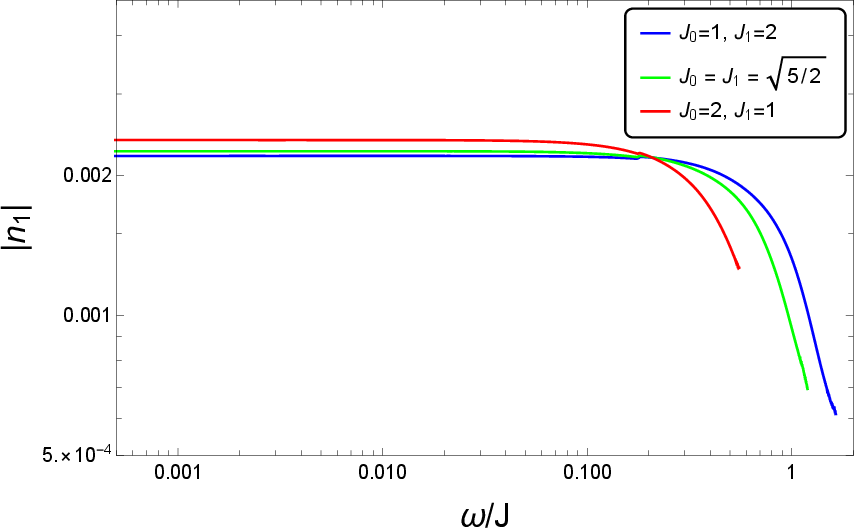}
	\caption{The absolute value of the off-diagonal elements $\mathcal{O}_{mn}$ of the particle number operator, $\mathcal{O} = \hat{n}_1$, as a function of the energy eigenvalue difference $\omega/J = (E_m - E_n)/J$, at a fixed average eigenenergy $(E_m + E_n)/2$ for three different coupling strengths $J_0$ and $J_1$. The ensemble average is taken over 1000 disorder realizations of the SYK chain model with $M = 4$ sites and $N = 6$ Majorana fermions per site. The average energy is fixed at $(E_m + E_n)/2NM = 0.02 \pm 0.001$, and a moving average in $\omega$ over the 800 nearest neighbors of the matrix elements is performed. The exponentially small values of the off-diagonal elements are in agreement with ETH. Additionally, the curves show a dependence on the ratio $J_1/J$ for larger $\omega/J$, indicating that the Thouless energy and the range of RMT-like behavior increase as $J_1/J$ increases, consistent with theoretical results for the diffusion constant of the model.}
	\label{fig:offdiagonalnumopvsden6m4real1000}
\end{figure}

For the hopping operator $\hat{h}_{13}$, the diagonal matrix elements are strictly zero due to symmetry, as discussed and reported for single realizations in Sec.~\ref{sec:SingleRealM4}. Consequently, their disorder-averaged values are also zero and in exact agreement with the corresponding microcanonical value.

\begin{figure}
	\includegraphics[width=\linewidth]{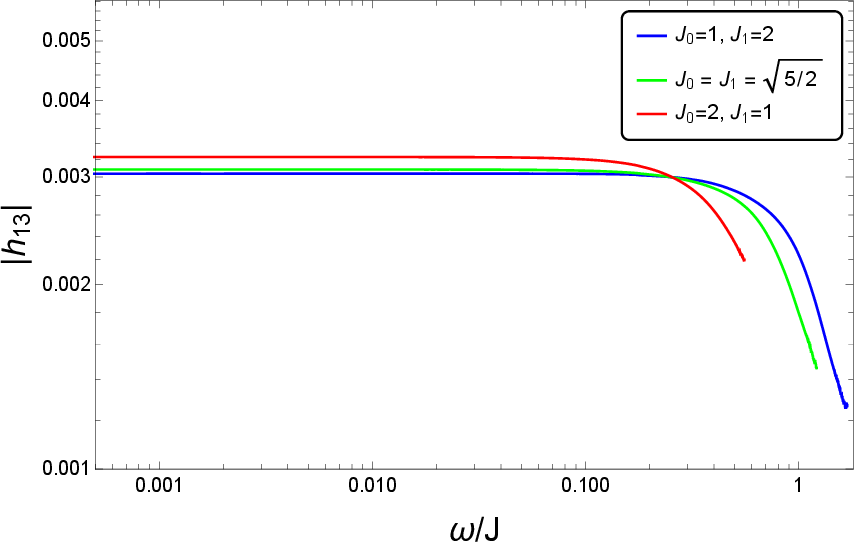}
	\caption{The absolute value of the off-diagonal elements $\mathcal{O}_{mn}$ of the hopping operator between sites 1 and 3, $\mathcal{O} = \hat{h}_{13}$, as a function of the energy eigenvalue difference $\omega/J = (E_m - E_n)/J$, at a fixed average eigenenergy $(E_m + E_n)/2$, for three different coupling strengths $J_0$ and $J_1$. The ensemble average is taken over 1000 disorder realizations of the SYK chain model with $M = 4$ sites and $N = 6$ Majorana fermions per site. The average energy is fixed at $(E_m + E_n)/2NM = 0.02 \pm 0.001$, and a moving average is performed in $\omega$ over 800 nearest neighbors of the matrix elements. The exponentially small values of the off-diagonal elements are consistent with ETH. Additionally, the curves exhibit a dependence on the ratio $J_1/J$ for larger $\omega/J$, indicating that the Thouless energy and the range of RMT-like behavior increase as $J_1/J$ increases, in accordance with theoretical results for the diffusion constant of the model.}
	\label{fig:offdiagonalhopopvsden6m4real1000}
\end{figure}

In contrast, the off-diagonal matrix elements are not constrained by symmetry and must be determined numerically. The ensemble-averaged absolute values of the off-diagonal matrix elements of the number operator $\hat{n}_1$ and the hopping operator $\hat{h}_{13}$ are computed and presented in Figs.~\ref{fig:offdiagonalnumopvsden6m4real1000} and \ref{fig:offdiagonalhopopvsden6m4real1000}, respectively.

As shown in these figures, the ETH predictions are satisfied for the off-diagonal matrix elements of both $\hat{n}_1$ and $\hat{h}_{13}$. The ensemble-averaged absolute values of the off-diagonal matrix elements are smaller than $e^{-\bar{S}/2}$ (see Tables~\ref{maxmeanm4n1} and~\ref{maxmeanm4h13} for the values of $e^{-\bar{S}/2}$). Thus, ETH is satisfied for all three coupling regimes considered.

Moreover, the behavior of the Thouless energy $E_T$, determined by the crossover from the $\omega$-independent regime (characteristic of RMT) to the $\omega$-dependent regime (non-RMT behavior), reveals an interesting trend. Specifically, as $J_1$ increases from $1$ (red curves) to $\sqrt{5/2}$ (green curves) and to $2$ (blue curves), while keeping $J = \sqrt{5}$ fixed, the value of $E_T$ increases. These findings are consistent with the analytic results for the diffusion constant of the SYK chain model reported in Ref.~\cite{Gu2017}, Eq.~\eqref{D2}, and the corresponding definitions of the Thouless energy, Eqs.~\eqref{ET} and \eqref{ET2}.

Therefore, I conclude that as the inter-site coupling $J_1$ increases, the range of $\omega$ over which the system exhibits RMT-like chaotic behavior becomes larger.

\section{Discussion and Conclusion} \label{con}

In this work, I have investigated thermalization via the eigenstate thermalization hypothesis (ETH) in the strongly correlated, non-local (1+1)-dimensional SYK chain model and its special case, the two-site SYK model. Using exact diagonalization, I examined two representative few-body operators: the local particle number operator acting on a single site and the non-local hopping operator acting between two sites. Both single realizations and ensemble averages over many disorder realizations were analyzed. The main goal was to verify whether the ETH conditions hold, both approximately and strictly, and to understand the mechanisms underlying thermalization in these systems.

The key findings are as follows. For individual disorder realizations, the diagonal elements of the observables in the energy eigenbasis fluctuate around the corresponding thermal values, and the off-diagonal elements are small but not exponentially suppressed. Consequently, the ETH is satisfied approximately in single realizations, allowing for thermalization with slightly larger fluctuations around equilibrium. In contrast, the ensemble-averaged theory satisfies the ETH conditions more precisely: the diagonal elements follow smooth energy-dependent curves, and the off-diagonal elements are exponentially small, fully consistent with the ETH ansatz.

An important conclusion is that thermalization occurs much faster than might have been inferred from earlier studies based on R'enyi entropies~\cite{Gu2017b}. The faster thermalization revealed by ETH analysis may have several explanations. One possibility is as follows. In the SYK chain model at strong coupling ($N \gg \beta J \gg 1$), the dominant light degrees of freedom are collective modes associated with the time-reparametrization field. These soft modes can lead to rapid equilibration of few-body observables, even when the system is spatially local and lacks all-to-all interactions. The heavy modes that Ref.~\cite{Gu2017b} identified through R'enyi entropy studies, which were claimed to induce slow thermalization, do not appear crucial for the ETH mechanism of thermalization.

Additional insight comes from Ref.\cite{Sohal2022}, which appeared after the first version of this work was posted on arXiv\cite{Halataei2021} and cited it. In that study, the authors show that thermalization in the SYK chain is state-dependent: states with sufficiently high effective temperature thermalize rapidly, even with respect to R'enyi entropies. Moreover, their analysis provides evidence that the previously reported subthermal behavior~\cite{Gu2017b} is a large-$N$ artifact, rather than an intrinsic feature of the model at finite $N$. Their results are consistent with and further support the conclusions drawn in the present work regarding the relatively fast thermalization observed through ETH analysis.

Additionally, I found that the Thouless energy $E_T$, which marks the crossover from random matrix theory (RMT)-like behavior to non-RMT behavior, increases with the strength of the inter-site coupling $J_1$. This observation matches theoretical expectations based on diffusion of energy in the SYK chain, where the diffusion constant grows with increasing $J_1$~\cite{Gu2017}. As the inter-site coupling strengthens, energy spreads more efficiently between sites, leading to an enlarged energy window over which RMT-like chaotic behavior persists.

These findings have broader implications. First, the spatial locality of the SYK chain model does not obstruct thermalization via ETH, suggesting that ETH holds beyond zero-dimensional models with all-to-all couplings. Second, from a holographic perspective, since the SYK chain is conjectured to be dual to an incoherent black hole, the validity of ETH for few-body operators indicates that black holes in the dual description also exhibit rapid thermalization.

Comparing with previous ETH studies in zero-dimensional SYK-like models~\cite{Sonner2017,Hunter2018}, this work extends the understanding to (1+1)-dimensional systems with spatial locality, demonstrating that thermalization mechanisms remain robust even when spatial structure is introduced. The results suggest that the dominant physics responsible for ETH and rapid relaxation is tied to collective low-energy dynamics.

Several future directions arise naturally from this work. While I focused on two conventional operators, it would be valuable to systematically examine whether more generic few-body operators also display ETH behavior in the SYK chain and its generalizations. Additionally, understanding how thermalization in ensemble-averaged theories can be rigorously derived from ETH at the level of individual realizations remains an open and important question. Finally, extensions of this analysis to higher-dimensional SYK models introduced in Refs.~\cite{Berkooz2017,Gu2017} would shed further light on the universality of ETH across strongly interacting quantum chaotic systems.

\bigskip

Overall, this study demonstrates that despite the absence of all-to-all interactions and despite earlier indications of slow entanglement growth, the SYK chain and two-site SYK models thermalize rapidly through the ETH mechanism, confirming the robustness of eigenstate thermalization in strongly interacting lattice systems.
	
	
\section{acknowledgments}
	I would like to thank Mohsen Alishahiha for especial inspiration, valuable discussions, and fruitful comments. I also thank Mark Srednicki, Manuel Vielma, Mohammad Hassan Vahidinia, Hamid Afshar, Seyed Hamed Aboutalebi, Julian Sonner, Davoud Nasr Esfahani, Abolhassan Vaezi, Mohammad Sadegh Vaezi, and Mahdieh Piranaghl for fruitful discussions. I especially acknowledge Junyu Lin for his  precious help with the code at an early stage. I would like to thank members of the IPM Turin Cloud Services (turin.ipm.ir) for delivering part of the computing power to run the simulations.


\bibliographystyle{unsrt}
\bibliography{../Bibliography/allrefs} 
	
\end{document}